\documentclass[11pt, a4paper]{article}

\usepackage{fullpage}
\usepackage{latexsym}
\usepackage{psfig}

\newcommand{\beq}{\begin{equation}}
\newcommand{\eeq}{\end{equation}}
\newcommand{\ba}{\begin{array}}
\newcommand{\ea}{\end{array}}
\newcommand{\bea}{\begin{eqnarray}}
\newcommand{\eea}{\end{eqnarray}}
\newcommand{\bc}{\begin{center}}
\newcommand{\ec}{\end{center}}
\newcommand{\bt}{\begin{table}}
\newcommand{\et}{\end{table}}
\newcommand{\eps}{\epsilon}
\newcommand{\la}[1]{\label{#1}}
\newcommand{\p}{\partial}
\newcommand{\bbox}{\rule{3mm}{3mm}}

\newcommand{\ds}{\displaystyle}
\newcommand{\no}{\noindent}
\newcommand{\pp}[2]{{\partial #1 \over \partial #2}}
\newcommand{\ppn}[3]{{\partial^{#1} #2 \over \partial #3^{#1}}}

\newcommand{\mbf}[1]{\mbox{\boldmath {$#1$}}}
\newcommand{\binomial}[2]{\left(\ba{c}#1 \\ #2 \ea\right)}
\newcommand{\rf}[1]{(\ref{#1})}
\newcommand{\beqno}{\begin{displaymath}}
\newcommand{\eeqno}{\end{displaymath}}
\newcommand{\dd}[2]{{\delta #1 \over \delta #2}}

\newcommand{\been}{\begin{enumerate}}
\newcommand{\een}{\end{enumerate}}
\renewcommand{\aa}[2]{\alpha_{#1}^{(#2)}(r)}

\newcommand{\since}[1]{\stackrel{\mbox {using \rf{#1}}}{=}}

\newcommand{\call}{{\cal L}}

\newlength{\myheight}
\newlength{\mylength}

\makeatletter
\let\c@table=\c@mytable
\makeatother

\makeatletter
\let\c@figure=\c@myfigure
\makeatother

\makeatletter
\let\c@equation=\c@myequation
\makeatother

\renewcommand{\theequation}{\arabic{section}.\arabic{equation}}

\newcommand{\pb}[1]{\parbox[c][\myheight]{\mylength}{\centerline{#1}}}
\newcounter{saveeqn}

\newcommand{\alpheqn}{\setcounter{saveeqn}{\value{equation}}
\stepcounter{saveeqn}\setcounter{equation}{0}
\renewcommand{\theequation}{\mbox{\arabic{section}.\arabic{saveeqn}
\alph{equation}}}}

\newcommand{\resetalpheqn}{\setcounter{equation}{\value{saveeqn}}
\renewcommand{\theequation}{\arabic{section}.\arabic{equation}}}

\newtheorem{lemma}{Lemma}[section]

\newtheorem{theo}{Theorem}[section]

\newtheorem{example}{Example}[section]

\begin{document}

\title{Canonical Variables for multiphase solutions of the KP equation}

\author{Bernard Deconinck\\
Mathematical Sciences Research Institute\\
1000 Centennial Drive\\
Berkeley Ca 94720\\
~\\
{\small \bf Keywords}:\\ {\small KP equation, Initial-value problem, 
Canonical variables},\\
{\small Quasiperiodic solutions, Complete integrability}
}

\maketitle

\begin{abstract}

The KP equation has a large family of quasiperiodic multiphase solutions. These
solutions can be expressed in terms of Riemann-theta functions. In this paper,
a finite-dimensional canonical Hamiltonian system depending on a finite number
of parameters is given for the description of each such solution. The
Hamiltonian systems are completely integrable in the sense of Liouville. In
effect, this provides a solution of the initial-value problem for the
theta-function solutions. Some consequences of this approach are discussed.

\end{abstract}


\section{Introduction}

In 1970, Kadomtsev and Petviashvili \cite{kadom} derived two equations as
generalizations of the Korteweg-de Vries (KdV) equation to two spatial
dimensions:

\renewcommand{\theequation}{\mbox{KP}}

\beq\la{kp}
\left(-4 u_t+6 u u_x+u_{xxx}\right)_x+3 \sigma^2 u_{yy}=0,
\eeq

\setcounter{equation}{0}
\renewcommand{\theequation}{\mbox{\arabic{equation}}}

\no where $\sigma^2=\pm 1$ and the subscripts denote differentiation. 

Depending on the physical situation, one derives the equation
either with $\sigma^2=-1$ or $\sigma^2=+1$. The resulting partial differential
equations are referred to as (KP1) and (KP2) respectively. For real-valued
solutions, the two equations have different physical meaning and different
properties \cite{kadom}. Nevertheless, for some purposes the sign of $\sigma^2$
is irrelevant and we equate $\sigma \equiv 1$. In this case, we simply refer to
``the KP equation'' or to ``(KP)''. This amounts to working with (KP2). If
necessary, (KP1) is obtained through a complex scaling of the $y$-variable.

The KP equation can be written as the compatibility condition of two linear
equations for an auxiliary wave function $\Psi$ \cite{dryuma, shabat}:

\alpheqn
\bea\la{linear1}
\sigma \Psi_y&=&\Psi_{xx}+u \Psi,\\\la{linear2}
\Psi_t&=&\Psi_{xxx}+\frac{3}{2} u \Psi_x+\frac{3}{4} \left(u_x+w\right) \Psi.
\eea
\resetalpheqn

Expressing the compatibility of \rf{linear1} and \rf{linear2},
$\Psi_{yt}\equiv \Psi_{ty}$, and eliminating $w$ results in \rf{kp}, if we
assume that a complete basis for the wave function $\Psi$ exists. Note that if
the KP solution is independent of $y$, the above Lax pair (\ref{linear1},
\ref{linear2}) reduces to the Lax pair for (KdV) \cite{ggkm} by simple
separation of variables of the wave function $\Psi$.

The two equations \rf{linear1}, \rf{linear2} are only two equations of an
infinite hierarchy of linear evolution equations for the wave function $\Psi$
with respect to {\em higher-order time variables} $t_n$ \cite{sato}. We refer
to this collection of linear flows as the {\em linear KP hierachy}:

\beq\la{lkphier}
\Psi_{t_k}=A_k \Psi,~~\mbox{for~}k=1,2,3,\ldots
\eeq

\no with $A_k$ a linear differential operator in $x$ of order $k$ with
(usually) nonconstant coefficients. We refer to the $t_k$, $k=1,2,3,4, \ldots$
as higher-order time variables. A consequence of our definition of the KP
hierarchy given in Section 3 is that $t_1$ can be identified with $x$. 
Furthermore, $y$ and $t$ are related to $t_2$ and $t_3$ respectively.  

By expressing the compatibility of these different linear flows,
$\Psi_{t_{k_1}t_{k_2}}=\Psi_{t_{k_2}t_{k_1}}$, and assuming the existence of a
complete basis for $\Psi$, we obtain an infinite number of nonlinear partial
differential equations for the evolution of $u$, $w$ (and other functions
referred to as potentials) with respect to the $t_k$ \cite{sato}, called the
{\em KP hierarchy}:

\beq\la{kphier}
\pp{A_{k_1}}{t_{k_2}}-\pp{A_{k_2}}{t_{k_1}}=\left[A_{k_2},A_{k_1}\right],
\eeq

\no where $\left[A,B\right]\equiv AB-BA$. The linear KP hierarchy \rf{lkphier}
and the KP hierarchy \rf{kphier} are the fundamental ingredients for the
methods presented in this paper.

A large family of quasiperiodic solutions of the KP equation was found by
Krichever \cite{krich1, kricheverintegration}. Each of these solutions has a
finite number of phases. They are given by

\alpheqn

\bea\la{reconstruction}
u&=&\tilde{u}+2 \p_x^2 \ln \Theta_g(\phi_1, \phi_2, \ldots, \phi_g |
\mbf{B}),\\\la{phases}
\phi_j&=&k_j x+l_j y+\omega_j t+\phi_{0j},~~\mbox{for~}j=1,2,\ldots, g,
\eea

\resetalpheqn

\no for some constants $\tilde{u}, k_j, l_j, \omega_j$ and $\phi_{0j}$, $j=1,
2,\ldots, g$. $\Theta_g$ is a {\em Riemann theta function} of {\em genus} g,
parametrized by a $g \times g$ {\em Riemann matrix} $B$:

\beq\la{thetafunction} \Theta_g(\mbf{\phi}| \mbf{B})\equiv \sum_{\mbf{m}\in
\mbf{Z}^g} \exp\left(\frac{1}{2} \mbf{m} \cdot \mbf{B} \cdot \mbf{m}+i \mbf{m}
\cdot \mbf{\phi}\right). \eeq

\no Here $\mbf{\phi}\equiv(\phi_1, \ldots, \phi_g)$. The vector $\mbf{m}$ runs
over all $g$-dimensional vectors with integer components. The Riemann matrix
$\mbf{B}$ is a symmetric $g \times g$ matrix with negative definite real part.
Whenever the matrix $\mbf{B}$ is obtained from a compact, connected Riemann
surface in a standard way \cite{dub}, \rf{reconstruction}
defines a solution of the KP equation \cite{krich1, kricheverintegration}. In
what follows, the dependence of the theta function $\Theta_g$ on the Riemann
matrix $\mbf{B}$ and the index $g$ denoting the number of phases will be
surpressed for notational simplicity. By construction, $\Theta(\mbf{\phi})$ is
periodic in each component of $\mbf{\phi}$. Hence the restriction to the linear
winding \rf{phases} makes \rf{reconstruction} by definition a quasiperiodic
function in $x$ or $y$ or $t$. A solution of the form \rf{reconstruction} is
said to have genus $g$. In the terminology of Krichever and Novikov
\cite{kricheverrank, krichrank}, all solutions of the form
\rf{reconstruction} have {\em rank 1}.

The problem addressed in this paper is to find $u(x,y,t)$ such that:

\beq\la{problem}
\left\{
\ba{l}
(-4 u_t+6 u u_x+u_{xxx})_x+3 \sigma^2 u_{yy}=0\\
u(x,y,0)=\mbox{rank 1, finite-genus solution of KP,}\\
\phantom{u(x,y,0)=}~\mbox{evaluated at~}t=0.
\ea
\right.
\eeq

\no The initial data are purposely not written in the form \rf{reconstruction}.
The problem is not only to determine the frequencies $\omega_j$, for
$j=1,2,\ldots, g$, but also to fix the other parameters  $g, \mbf{B},
\tilde{u}, k_j, l_j, \phi_{0j}$, for $j=1,2,\ldots, g$: the acceptable class
of initial data consists of all finite-genus solutions of the KP equations
evaluated at a fixed time, with an unspecified but finite number of phases.

A solution to this problem was offered in \cite{ds1}, where a seven-step
algorithm was presented.  This algorithm was a mixture of new ideas and
known work by Krichever \cite{kricheverchap, krich1,
kricheverintegration} and Previato \cite{previato}. The main idea of
the algorithm is to provide the ingredients required for Krichever's inverse
procedure for the reconstruction of a finite-genus solution of the KP equation
\cite{krich1, kricheverintegration}, {\em i.e.} a compact connected Riemann
surface and a divisor on this surface.

In the present paper, a different algorithm to solve the problem \rf{problem}
is presented. This algorithm shares some steps with that in \cite{ds1}.
However, in contrast to the first algorithm, the second algorithm does not
work towards Krichever's inverse procedure \cite{krich1,
kricheverintegration}.  The main idea here is to examine the structure of a
set of ordinary differential equations obtained in step 5 of \cite{ds1}. 
In this paper, we show the following:

\begin{itemize}

\item The rank 1, finite-genus solutions of the KP equation are governed by a
system of ordinary differential equations. This system is constructed
explicitly. 

\item This system of ordinary differential equations is Lagrangian. 

\item With some care, the Lagrangian equations are written as a Hamiltonian
system of ordinary differential equations in $x$.

\item This Hamiltonian system of ordinary differential equations is completely
integrable in the sense of Liouville \cite{arnold}. A complete set of
conserved quantities in involution under the Poisson bracket is explicitly
constructed. 

\item From these conserved quantities, one easily constructs completely
integrable Hamiltonian systems of ordinary differential equations describing
the evolution of the given initial condition of \rf{problem} under any of the
higher-order time variables $t_k$, including $k=1,2,3$. This provides a
solution of \rf{problem}.

\end{itemize}

In the next section, it is shown how the information listed above is used in
an algorithm to solve problem \rf{problem}. As with the algorithm in
\cite{ds1}, most of the steps of the algorithm in this paper are due to
others. The work of Bogoyavlenskii and Novikov \cite{bogoyavlenskii} provided
the main inspiration: the algorithm presented here is a generalization to the
KP equation of their approach to solve problem \rf{problem} for the KdV
equation. The work by Gel'fand and Dikii \cite{gd1}, Veselov \cite{veselov},
Adler \cite{adler} and Dikii \cite{dickey} was used to execute some of the
steps of the algorithm. Although all of the above authors have considered
parts of the problem considered here, to the best of our knowledge a complete
solution of problem \rf{problem} using ordinary differential equations to
solve the posed initial-value problem was never given.

The algorithm presented here offers an alternative approach to that in
\cite{ds1}. There are however some natural consequences of the new approach
that do not appear from the approach in \cite{ds1}. These include

\begin{itemize}

\item {\bf Canonical variables} for the rank 1, finite-genus solutions of the
KP equation. Any rank 1, finite-genus solution of the KP equation satisfies a
Hamiltonian system of ordinary differential equations. The Poisson bracket on
the phase space of this Hamiltonian system is the canonical Poisson bracket,
resulting in a description of the rank 1, finite-genus solutions of the KP
equation in terms of canonical (Darboux) variables.

\item {\bf Conserved Quantities} for rank 1, finite-genus solutions
\rf{reconstruction} of the KP equation. The Hamiltonian system of ordinary
differential equations is completely integrable in the sense of Liouville. A
sufficient number of conserved quantities is constructed explicitly. These
conserved quantities are mutually in involution under the canonical Poisson
bracket.

\item {\bf Parameter count} of the theta-function solutions \rf{reconstruction}
of the KP equation. It is known \cite{dub} that a generic\footnote{for a
precise definition, see section \ref{sec:algorithm}} solution of genus $g$ of
the KP equation is characterized by $4g+1$ independent parameters. We reconfirm
this result, and extend it, by providing a parameter count for nongeneric
solutions of the KP equation as well. Furthermore, the parameters naturally
divide in two classes:  parameters with ``dynamical significance'' and other
parameters. The dynamically significant parameters are the initial conditions
of the Hamiltonian system of ordinary differential equations describing the
solution. The other parameters foliate the phase space of the Hamiltonian
system.

\item {\bf Minimal characterization of the initial data}. The approach
presented here demonstrates that a finite-genus solution $u(x,y,t)$ of the KP
equation is completely determined by a one-dimensional slice of the initial
condition $u(x,y=0,t=0)$. In other words, it suffices to specify the initial
data of \rf{problem} at a single $y$-value, say at $y=0$.

\end{itemize}

Krichever \cite{krich3} proposed another method to solve an initial-value
problem for the KP2 equation with initial data that are spatially periodic (in
both $x$ and $y$). Krichever's method is not restricted to initial data of
finite genus, hence it is in that sense more general than the algorithm
presented here. On the other hand, the methods of this paper require no
restriction to periodic initial data.  



\section{Overview of the algorithm}\la{sec:algorithm}

A solution for problem \rf{problem} is obtained using a seven-step algorithm.
In this section, an overview of this algorithm is given, along with references
to earlier work.

\begin{enumerate}

\item {\bf Determine the genus of the initial data} \cite{ds1} Let us rewrite
\rf{phases} in the form

\beq\la{newphases}
\phi_j=\mbf{\kappa_j} \cdot \mbf{x}+\omega_j t+\phi_{0j}, ~~~~j=1,2,\ldots, g,
\eeq

\no with $\mbf{\kappa_j}=(k_j,l_j)$ and $\mbf{x}=(x,y)$. If all wave vectors
$\mbf{\kappa_j}$ are incommensurable, {\em i.e.,} if there is no relationship

\beq\la{commensurable}
\sum_{i=1}^g n_i \mbf{\kappa_i}=0
\eeq

for integers $n_i$ not all zero, then a two-dimensional Fourier transform of
the initial data resolves the vectors $\mbf{\kappa_j}$. Because the initial
data contain only a finite number of phases, the Fourier transform is
necessarily discrete; {\em i.e.,} it consists of isolated spikes. Since the
condition \rf{commensurable} almost never holds, we can almost always find the
genus of the initial condition by counting the number of spikes in the Fourier
transform, modulo harmonics. 

If condition \rf{commensurable} holds, then the prescribed method finds only a
lower bound on the genus of the initial data. The method fails especially
dramatically in one important special case: if the initial data are spatially
periodic, then \rf{commensurable} holds automatically for any two wave vectors
$\mbf{\kappa_i}$ and $\mbf{\kappa_j}$. The lower bound obtained in this case
for the number of phases is 1. This problem was already pointed out in
\cite{ds1}. If the initial data are spatially periodic, it is most convenient
to impose that the genus of the initial data also be given, as part of the
initial data.

The method of Fourier transform to determine the genus of the initial condition
has been used in \cite{hmss, currysegur}. 

\item {\bf Determine two stationary flows of the KP hierarchy: find $(r,n)$}

Mulase \cite{mulase} and later Shiota \cite{shiota} showed that a rank 1,
finite-genus solution of the KP equation \rf{reconstruction} is a simultaneous
solution to all flows of the KP hierarchy by using \rf{reconstruction} with 

\beq\la{phaseshier}
\phi_j=\sum_{i=1}^\infty k_{j,i} t_i,
\eeq

\no for $j=1,2,\ldots, g$, instead of \rf{phases}.  Mulase and Shiota
demonstrated that the corresponding rank 1, finite-genus solutions are
stationary with respect to all but a finite number of the higher-order times
in the KP hierarchy.  A rank 1, finite-genus solution of the KP
equation is said to be stationary with respect to $t_k$ if

\beq\la{stat}
\sum_{i=1}^k d_{i} \pp{u}{t_i}=0,
\eeq

\no with all the $d_{i}$ constant and $d_{k}=1$.  

The algorithm presented here requires the knowledge of two independent
higher-order times of the KP hierarchy $t_r$ and $t_n$, such that $u$ is
stationary with respect to both $t_r$ and $t_n$. First, $r$ is the minimal $k$
for which \rf{stat} holds for $k=r$. For this $r$, $n$ corresponds to the
lowest-order higher-order time $t_n$, such that the $t_n$-flow is independent
of the $t_r$-flow and \rf{stat} holds for $k=n$.

In \cite{ds1}, a recipe was presented to find $(r,n)$, given the genus $g$ of
the initial data. Actually, a finite number of pairs $(r,n)$ is determined for
any given $g$. As we will see in step 4, each one of the resulting pairs
$(r,n)$ gives rise to a set of ordinary differential equations, one of which
the initial condition necessarily satisfies. The pairs $(r,n)$ for which the
initial condition does not satisfy the differential equations need to be
rejected. Hence, only at step 4 do we nail down a pair of stationary flows of
the KP hierarchy for the given initial data. Here, at step 2, the numbers of
pairs $(r,n)$ is reduced to a finite number. 

For initial data with $g$ phases, the following constraints on $(r,n)$ are
known \cite{ds1}:

\begin{itemize}

\item All values of $r$ with $2 \leq r \leq g+1$ are allowed. 

\item For each $r$, let $n_j(r)$ be the $j$-th integer greater than $r$ that is
coprime with $r$. The lowest $(g-r+2)$ of these integers are possible values of
$n$. 

\item Exclude from the list of pairs $(r,n)$ obtained above the values of $n$
for which \linebreak $(r-1)(n-1)/2<g$.

\item The remaining pairs $(r,n)$ are all possible for genus $g$.

\end{itemize}

\newpage

\no {\bf Remarks}\la{remrem}

\begin{enumerate}

\item When $r=2$, the only possibility for $n$ is $n=2g+1$. This is
the case corresponding to one-dimensional solutions.

\item A solution of the KP equation of genus $g$ is called {\em generic} if
the first $g$ vectors $\mbf{k_i}=(k_{1,i},k_{2,i}, \ldots, k_{g,i})$ are
linearly independent. For a generic solution of genus $g$ of the KP equation,
$(r,n)=(g+1, g+2)$. 

\item We have the following confusing situation: to find a generic genus $g$
KP solution, we need $(r,n)=(g+1,g+2)$. This choice leads to a Hamiltonian
system of ordinary differential equations (in step 5). However, a generic
solution of genus $g$ is not a typical solution of this Hamiltonian system
({\em i.e.,} it does not  depend on a maximal number of parameters for this
system). Typical solutions of the Hamiltonian system are nongeneric solutions
of higher genus. Since these higher-genus solutions depend on more parameters,
one must search carefully to find the generic genus $g$ solutions among them.
This is further discussed in Sections \ref{sec:reductions} and
\ref{sec:parameters}.

\end{enumerate}

\item {\bf Impose the $r$-reduction}

For a given value of $r$, obtained from step 2, we impose on the KP hierarchy
\rf{kphier} the reduction that the KP solution is independent of $t_r$. Hence,
the coefficients of $A_k$, for all k are independent of $t_r$. Following
Gel'fand and Dikii \cite{gd1}, Adler \cite{adler} and Strampp and Oevel
\cite{strampp}, this allows us to rewrite the {\em $r$-reduced KP hierarchy}
as an infinite ($r\times r$ matrix) hierarchy of partial differential
equations, each with one space dimension, and all of them mutually commuting.
In other words, \rf{kphier} is replaced by a hierarchy of the form

\beq\la{11hier}
\pp{B}{t_k}=\left[B_k,B\right], ~~~r~\mbox{does not divide}~k.
\eeq

The matrices $B$ and $B_k$ contain $r-1$ unknown functions. The higher-order
time variables of the hierarchy in \rf{11hier} are inherited from the KP
hierarchy \rf{kphier}. Only $t_r$ and the higher-order times of the form
$t_{(ir)}$ for integer $i$ do not appear any more. In particular $t_1\equiv
x$. Each equation of the hierarchy \rf{11hier} is Hamiltonian, as is shown in
Section \ref{sec:rred}, where the details of the $r$-reduction are given. 

\item {\bf Impose the $n$-reduction}

After imposing stationarity of the KP solution with respect to $t_r$, we now
impose stationarity of the KP solution with respect to $t_n$ as well. Imposing
the $n$-reduction in addition to the $r$-reduction leads to the {\em
$(r,n)$-reduced KP equation}. The $(r,n)$-reduced KP equation is a system of
$(r-1)$ ordinary differential equations in $x$ for $(r-1)$ unknown functions
$\mbf{u}$. 

Again, following Gel'fand and Dikii \cite{gd1}, Adler \cite{adler} and Strampp
and Oevel \cite{strampp}, we write the $(r,n)$-reduced KP equation in
Lagrangian form:

\beq\la{lagode}
\dd{\call}{\mbf{u}}=\mbf{0},
\eeq

\no where $\delta{\call}/\delta{\mbf{u}}$ denotes the variational derivative
of $\call$ with respect to a certain vector function $\mbf{u}=(f_1,
f_2, \ldots, f_{r-1})$, which is explicitly determined in terms of the
solution of (KP):

\beq\la{vardervec} \dd{\call}{\mbf{u}}\equiv\left(\dd{\call}{f_1},
\dd{\call}{f_2}, \ldots, \dd{\call}{f_{r-1}}\right)^T \eeq

\no and for any function $f$, the {\em variational derivative} of ${\cal L}$
with respect to $f$ is defined as

\beq\la{varder}
\dd{}{f}\call(u,u_x, u_{xx}, \ldots)\equiv \sum_{k\geq 0} (-1)^k \ppn{k}{}{x}
\pp{\call}{f^{(k)}}.
\eeq

\no Here, $f^{(k)}$ denotes the $k$-th derivative of $f$ with respect to $x$.

Equations \rf{lagode} are a set of ordinary differential equations that the
initial condition needs to satisfy. This constitutes a test on the validity of
the pair $(r,n)$, chosen after step 2.

The details of imposing the $n$-reduction in addition to the $r$-reduction are
found in Section \ref{sec:nred}

\item {\bf The Ostrogradskii transformation, canonical variables and the
Hamiltonian system}

In Section \ref{sec:ostro}, the Lagrangian system of ordinary differential
equations in $x$ is transformed to a Hamiltonian system of ordinary
differential equations in $x$ with canonical variables. Since the Lagrangian
${\cal L}$ depends on more than the first derivatives of $\mbf{u}$, an
extension of the Legendre transformation is needed. This is the {\em
Ostrogradskii transformation} \cite{ostro, whittaker}, defined in Section
\ref{sec:ostro}. It defines {\em canonical variables} $\mbf{q}$ and $\mbf{p}$
in terms of the Lagrangian variables $\mbf{u}$:

\beq\la{ostrosimple}
\mbf{q}=\mbf{q}(\mbf{u}, \mbf{u}_x, \mbf{u}_{xx}, \ldots), ~~~
\mbf{p}=\mbf{p}(\mbf{u}, \mbf{u}_x, \mbf{u}_{xx}, \ldots).
\eeq

The Lagrangian ${\cal L}$ is called {\em nonsingular} \cite{krupkova, dfn2} if
the Ostrogradskii transformation is invertible, {\em i.e.,} if the
transformation \rf{ostrosimple} can be solved for the Lagrangian variables
$\mbf{u}$ and their derivatives in terms of the canonical variables $\mbf{q}$
and $\mbf{p}$. If the Lagrangian is nonsingular, the Euler-Lagrange equations
corresponding to the Lagrangian ${\cal L}$ are equivalent to the Hamiltonian
system 

\beq\la{hamsyssimple}
\pp{\mbf{q}}{x}=\pp{H}{\mbf{p}},~~~ \pp{\mbf{p}}{x}=-\pp{H}{\mbf{q}}
\eeq

\no where the Hamiltonian $H$ is determined explicitly in terms of the
Lagrangian.

If both $r$ and $n$ are odd, the Lagrangian ${\cal L}$ is shown to be singular
in Section \ref{sec:sing}. Nevertheless, the dynamics in terms of the
Lagrangian variables is still well-posed, as shown by Veselov \cite{veselov}.
In Section \ref{sec:sing}, the singular Lagrangians are further investigated.
We indicate how one might be able to avoid dealing with singular Lagrangians:
a simple invertible transformation on the Lagrangian variables should be able
to transform the singular Lagrangian into a nonsingular one. Otherwise, one can
always resort to the more general methods of Krupkova \cite{krupkova} or to the
theory of constrained Hamiltonian systems \cite{dirac}.

\item {\bf Complete integrability of the Hamiltonian system}

The Hamiltonian system \rf{hamsyssimple} is shown to be {\em completely
integrable in the sense of Liouville} in Section \ref{sec:comp}. If the
dimension of the vectors $\mbf{q}$ and $\mbf{p}$ is $N$, the Hamiltonian system
is $2N$-dimensional. A set of $N$ functionally independent conserved quantities
$T_{k}$ is constructed. Generalizing the work of Bogoyavlenskii and Novikov
\cite{bogoyavlenskii}, these conserved quantities are
shown to be mutually {\em in involution}, {\em i.e.,}

\beq\la{involsimple}
\left\{T_{k},T_{l} \right\}\equiv 0,
\eeq

\no where $\{f,g\}$ denotes the {\em Poisson bracket} of the functions $f$
and $g$:

\beq\la{pb}
\left\{f,g\right\}\equiv \pp{f}{\mbf{q}} \pp{g}{\mbf{p}}-
\pp{f}{\mbf{p}} \pp{g}{\mbf{q}}.
\eeq

A consequence of proving the involutivity of the conserved quantities $T_{k}$
is that $T_k=-H_k$, where $H_k$ is the Hamiltonian describing the evolution of
the canonical variables along the higher-order time variable $t_k$:

\beq\la{hamsysksimple}
\pp{\mbf{q}}{t_k}=\pp{H_k}{\mbf{p}},~~~ \pp{\mbf{p}}{t_k}=-\pp{H_k}{\mbf{q}}.
\eeq

The canonical variables are related to the dependent variable $u$ of the KP
equation. Hence, we have constructed a set of ordinary differential Hamiltonian
systems, each one of which describes the evolution of a rank 1, finite-genus
solution of the KP equation according to a different higher-order time
variable. Since all these Hamiltonian systems are $2N$-dimensional and share a
common set of $N$ functionally independent conserved quantities $T_k$, mutually
in involution, they are all completely integrable in the sense of Liouville. 

\item {\bf Solve the Hamiltonian system; reconstruct the solution of the KP
equation}

The final step of the algorithm is to integrate explicitly the Hamiltonian
systems obtained in the previous step. From Liouville's theorem \cite{arnold}
it is known that the Hamiltonian equations of motion can be solved in terms of
quadratures. 

This last step is not executed in this paper. For the KdV equation
it can be found in \cite{dickey}. Some partial results for the KP equation are
also discussed there.

\end{enumerate}

\section{The KP hierarchy}\la{sec:hier}

In this section, the KP hierarchy is redefined, using the terminology of
Gel'fand and Dikii \cite{gd1}, Adler \cite{adler} and others. More
specifically, the notation of Strampp and Oevel \cite{strampp} is used. 

Consider the pseudo-differential operator

\beq\la{pseudo1}
L=\p+u_2\p^{-1}+u_3 \p^{-2}+u_4 \p^{-3}+\ldots=\sum_{j=-\infty}^{1}u_{1-j}\p^j,
\eeq

\no with $u_0\equiv 1$, $u_1\equiv 0$. We have used the notation $\p=\p_x$. The
coefficients $u_j$ can be functions of $x$. The $u_j$ are referred to as {\em
potentials}. This term is also used for any other set of functions, related to
the $u_j$ by an invertible transformation. 

\vspace*{12pt}
\no {\bf Remark}\la{u1rem}
In order to compare with the results in \cite{ds1}, we need $u_1\neq 0$, but
constant, extending the definition of the pseudo-differential operator $L$.
Although this changes some of the formulas in this section, the added results for
the KP equation are minor. In \cite{dickey} and \cite{decthesis}, it is shown that
this amounts to assigning a fixed value to the constant $\tilde{u}$ in
\rf{reconstruction}. In the remainder of the paper, we assume $u_1\equiv 0$, unless
stated otherwise.  
\vspace*{12pt}

The action of the operator $\p^j$ is defined by the generalized Leibniz rule: 

\beq\la{leibniz}
\p^j f=\sum_{i=0}^\infty \binomial{j}{i} f^{(i)} \p^{j-i},
\eeq

\no where $f$ is a function, $f^{(i)}$ is its $i$-th derivative with respect
to $x$, and the binomial coefficients are defined as

\beq\la{binomial}
\binomial{j}{i}=\frac{j(j-1)\cdots (j-i+1)}{i!}, ~~\mbox{for}~i>0,
~~\mbox{and}~\binomial{j}{0}=1.
\eeq

\no Note that this definition makes sense for negative integers $j$. For
non-negative integers $j$, \rf{leibniz} is a finite sum. Otherwise,
\rf{leibniz} results in an infinite series. 

Next, consider positive integer powers of the pseudo-differential operator $L$:

\bea\nonumber
L^r&=&\left(\p+u_2\p^{-1}+u_3\p^{-2}+u_4 \p^{-3}+\ldots\right)^r\\\la{def}
&=&\sum_{j=-\infty}^r \alpha_j(r)\p^j=\sum_{j=-\infty}^r \p^j \beta_j(r).
\eea

\no The last two equalities define the functions $\alpha_j(r)$ and
$\beta_j(r)$, for $j \leq r$, $r>0$. These are in general functions of $x$.
One has 
\beq\la{initialization}
\alpha_1(1)=1, \alpha_0(1)=0, \alpha_j(1)=u_{j+3}, ~~~~\mbox{for}~~j=-1, -2, -3,
\ldots,
\eeq
\no and
\beq \alpha_r(r)=1, ~~\beta_r(r)=1, ~~~\mbox{and}~~~\alpha_{r-1}(r)=0,~~
\beta_{r-1}(r)=0. \eeq

\no Clearly the functions $\alpha_{j}(r)$ and $\beta_j(r)$ are related. Using
\rf{leibniz}, we get from \rf{def} that

\beq\la{triangular}
\alpha_j(r)=\sum_{k=0}^{r-j}\binomial{j+k}{k} \beta_{j+k}^{(k)}(r).
\eeq

\no This triangular system can be solved to obtain the functions $\beta_j(r)$ in
terms of the functions $\alpha_j(r)$, if so desired. Note in particular that
$\alpha_{-1}(r)=\beta_{-1}(r)$, since the binomial coefficient \rf{binomial}
vanishes for positive $j$ less than $i$.

The functions $\alpha_j(r)$ can be determined explicitly in terms of the
potentials $(u_2, u_3, \ldots)$. A convenient way to do this is to use a
recursion relationship obtained from $L^r=LL^{r-1}$:
\beq\la{recursion}
\alpha_j(r)=\alpha_{j-1}(r-1)+\pp{}{x}\alpha_j(r-1)+u_{r-j}+\sum_{k=j-r+3}^{-1}
u_{1-k} \sum_{m=j}^{k+r-3}\binomial{k}{m-j}\alpha_{m-k}^{(m-j)}(r-1).
\eeq

\no It is possible to obtain an explicit formula which expresses $\alpha_j(r)$ in
terms of only the potentials $(u_2, u_3 \ldots)$ \cite{decthesis}, but such a
formula is not as practical as the recursion relationship \rf{recursion}.
However, the following result will be used. It extends a result of Date $et~al$
\cite{date}:

\beq\la{linearpart}
\alpha_j(r)=\sum_{k=1}^{r-j-1}\binomial{r}{k} \ppn{k-1}{}{x}
u_{r-j-k+1}+\hat{\alpha}_j(r),
\eeq

\no and $\hat{\alpha}_j(r)$ is a differential polynomial in $(u_2,u_3, \ldots,
u_{r-j-2})$ containing only nonlinear terms. This follows easily from
\rf{recursion}. 

The differential part (including the purely multiplicative term) of the operator
$L^r$ is denoted by $L^r_+$:
\beq\la{positive}
L^r_+=\sum_{j=0}^r \alpha_j(r)\p^j=\sum_{j=0}^r \p^j \beta_j(r).
\eeq

\no Observe from \rf{triangular} that the purely differential
part of $L^r$ is independent of the representation \rf{def} used for $L^r$. This
is also true for 

\beq\la{negative}
L^r_-=L^r-L^r_+=\sum_{j=-\infty}^{-1} \alpha_j(r)\p^j=\sum_{j=-\infty}^{-1} \p^j
\beta_j(r).
\eeq

Having introduced the above notation, the KP hierarchy is expressed quite easily.
Consider the linear evolution equation for the wave function $\Psi$

\beq\la{laxpsi}
\pp{\Psi}{t_r}=L^r_+ \Psi, ~~~~\mbox{for}~r=1,2,\ldots.
\eeq

\no This is the {\em linear KP hierarchy} ({\em cf.} \rf{lkphier}). For $r=1$,
this equation given $\Psi_{t_1}=\Psi_x$, hence the identification $t_1\equiv x$.
Assuming completeness of states, the {\em KP hierarchy} is obtained from the
compatibility of the equations in \rf{laxpsi} ({\em cf.} \rf{kphier}):

\beq\la{laxkp}
\frac{\p^2 \Psi}{\p_{t_{r_1}} \p_{t_{r2}}}=
\frac{\p^2 \Psi}{\p_{t_{r_2}} \p_{t_{r1}}}\Rightarrow
\pp{L_+^{r_1}}{t_{r_2}}-\pp{L_+^{r_2}}{t_{r_1}}=\left[L_+^{r_2},L_+^{r_1}\right].
\eeq

\no These equations determine how the potentials depend on the higher-order
time variables $t_r$ so that the equations \rf{laxpsi} are compatible.  Again
assuming completeness of states, equations \rf{laxkp} can also be obtained
from the compatibility of the following sequence of Lax-like equations 

\beq\la{lax}
\pp{L}{t_r}=\left[L_+^r,L\right]=\left[L,L_-^r\right], ~~r\geq 1.
\eeq

\no The last equality is a consequence of $L^r=L^r_++L_-^r$.

Introducing the KP hierarchy as in \rf{lax} is equivalent to the approach used in
\cite{ds1}. Below, we use that \rf{laxpsi} is essentially equivalent to \rf{lax}.
Our approach consists mainly of rewriting \rf{lax} and its reductions. 

Each time we increase $r$ by one, another potential appears in $L^r_+=L^r_+(u_2,
u_3, \ldots, u_r)$. Furthermore, $u_r$ appears only in $\alpha_0(r)$:
$\alpha_{0}(r)=r u_r+\tilde{\alpha}_0(r;u_2, u_3, \ldots, u_{r-1})$, as is seen
from \rf{recursion}. As a consequence, there is a one-to-one correspondence
between the potentials $u, w_1, w_2, \ldots$ appearing in the KP hierarchy as it
is defined in \cite{ds1} and the set of potentials $u_2, u_3, u_4,\ldots$
appearing in \rf{pseudo1}. 

As an example, consider equations (\ref{linear1}, \ref{linear2}). These are
contained in the formulation of the KP hierarchy given here: writing out
\rf{laxpsi} for $r=2$ and $r=3$ and equating coefficients with \rf{linear1} and
\rf{linear2} respectively gives $u_2=u/2$ and $u_3=w/4-u_x/4$.

The explicit form of the Lax equations \rf{lax} is needed later on. We have
\cite{strampp}

\beq\la{laxexp}
\pp{u_i}{t_r}=\sum_{j=1}^{i} M_{i,j}\beta_{-j}(r), ~~~~i=0,1,2, \ldots,
\eeq

\no where the differential operator $M_{i,j}$ is given by 

\beq\la{opera}
M_{i,j}=\sum_{k=0}^{i-j}\left(\binomial{1-j-k}{i-j-k}u_k
\p^{i-j-k}-\binomial{-j}{i-j-k}\p^{i-j-k}u_k\right).
\eeq

\no Here and in what follows, the contribution of a sum is assumed to be zero if
its upper limit is less than its lower limit, as happens in \rf{opera} when $i=0$.
Note that this immediately gives $\p u_0/\p t_r=0$ and $\p u_1/\p t_r=0$, for all
$r$, as expected. Furthermore, $M_{i,i}=0, M_{i,i-1}=\p$. The differential
equations \rf{laxexp} determine a first-order system for the $t_r$ evolution of
the infinite-dimensional vector of potentials $(u_2, u_3, u_4, \ldots)$.

\section{Impose the $r$-reduction}\la{sec:rred}

Next we obtain a closed first-order system of partial differential equations for
finitely many of the potentials, by imposing an {\em $r$-reduction}. This is the
first reduction step in our scheme towards our goal of finding a set of ordinary
differential equations describing the rank 1, finite-genus solutions of (KP). 

The $r$-reduction of the operator $L$ is obtained by imposing that the $r$-th power
of $L$ is purely differential:

\beq\la{rred}
L^r=L_+^r ~~\mbox{or}~~L^r_-=0~\Rightarrow~\beta_k(r)\equiv
0~\mbox{for}~k<0~\Rightarrow~\alpha_k(r)\equiv 0~\mbox{for}~k<0.
\eeq

\no Notice that the $r$-reduction implies immediately that all potentials are
independent of $t_r$, from \rf{laxexp} or \rf{lax}. The $r$-reduction determines
the potentials $u_{r+1}, u_{r+2}, u_{r+3}, \ldots$ as differential polynomials of
the potentials $u_2, u_3, \ldots, u_{r}$. This is a consequence of the triangular
structure of the system relating the potentials $(u_2, u_3, u_4, \ldots)$ to the
potentials $(\alpha_{r-2}(r), \alpha_{r-3}(r), \ldots)$.

\vspace*{12pt}
\no {\bf Remark}\la{p:rem2}
If we impose an $r$-reduction, for some positive integer number $r$, then we have
automatically achieved an $rk$ reduction, for any positive integer $k$. If $L^r$ is
purely differential, then so is $L^{rk}=(L^r)^k$. 
\vspace*{12pt}

Under $r$-reduction, the infinite system of evolution equations in \rf{laxexp}
reduces to a finite number of equations for the independent potentials $(u_2,
u_3, \ldots, u_r)$. We write this finite system in Hamiltonian form.  First,
we write the system in matrix-operator form. The matrices involved are now
finite-dimensional, as there are only $r-1$ independent potentials.

For a given $n$, define the $(r-1)$-dimensional vectors
\beq\la{capu}
\mbf{U}(r)=\left(
\ba{c}
u_2\\u_3\\u_4\\\vdots\\u_{r}
\ea
\right), ~~~
\mbf{\beta}(r,n)=\left(
\ba{c}
\beta_{-1}(n)\\\beta_{-2}(n)\\\beta_{-3}(n)\\\vdots\\\beta_{-r+1}(n)
\ea
\right),
\eeq

\no and the operator-valued matrix
\beq\la{opermat}
\mbf{M}(r)=\left(
\ba{ccccc}
M_{2,1}&0&0&\cdots&0\\
M_{3,1}&M_{3,2}&0&\cdots&0\\
\vdots&\vdots&\vdots&\ddots&\vdots\\
M_{r-1,1}&M_{r-1,2}&M_{r-1,3}&\cdots&0\\
M_{r,1}&M_{r,2}&M_{r,3}&\cdots&M_{r,r-1}
\ea
\right),
\eeq

\no with the operators $M_{i,j}$ defined by \rf{opera}. Then under $r$-reduction
\rf{laxexp} can be written as 
\beq\la{explicitlaxrred}
\pp{\mbf{U}(r)}{t_n}=\mbf{M}(r)\mbf{\beta}(r,n).
\eeq

In order to write the system of equations \rf{explicitlaxrred} in Hamiltonian
form, we first introduce new coordinates on the phase space of the system.
Define
\beq\la{capalpha}
\mbf{\alpha}(r)=\left(
\ba{c}
\alpha_0(r)\\\alpha_{1}(r)\\\vdots\\\alpha_{r-2}(r)
\ea
\right).
\eeq

\no The Jacobian matrix of the transformation from the coordinates
$\mbf{U}(r)\rightarrow \mbf{\alpha}(r)$ is the Fr\'{e}chet derivative of the
transformation (which depends also on the spatial
derivatives of the original coordinates, see for instance \rf{recursion}). The
Jacobian for such a transformation is an operator-valued matrix, whose action
on an arbitrary vector of functions $\mbf{v}=(v_2, v_3, \ldots, v_n)^T$ is
given by
\bea\nonumber
\mbf{D}(r) \mbf{v}&=&\pp{\mbf{\alpha}(r)}{\mbf{U}(r)} \mbf{v}\\\la{jacobian}
&=&\left.\pp{}{\eps}\left(
\ba{c}
\alpha_0(r)(u_2+\eps v_2, u_3+\eps v_3, \ldots, u_r+\eps v_r)\\
\alpha_1(r)(u_2+\eps v_2, u_3+\eps v_3, \ldots, u_r+\eps v_r)\\
\vdots\\
\alpha_{r-2}(r)(u_2+\eps v_2, u_3+\eps v_3, \ldots, u_r+\eps v_r)\\
\ea
\right)\right|_{\eps=0}.
\eea

\no This Jacobian matrix is upper-triangular. This is a direct consequence of
the triangular structure of the system relating the potentials $u_j$ to the
potentials $\alpha_j(r)$, $j=2,3,\ldots, r$. 

We rewrite the equations \rf{explicitlaxrred} in terms of the coordinates
$\mbf{\alpha}(r)$:
\beq\la{first}
\pp{\mbf{\alpha}(r)}{t_n}=\pp{\mbf{\alpha}(r)}{\mbf{U}(r)}
\pp{\mbf{U}(r)}{t_n}=\mbf{D}(r) \mbf{M}(r)
\mbf{\beta}(r,n)=\mbf{J}(r) \mbf{\beta}(r,n), 
\eeq

\no where we introduce the operator-valued matrix $\mbf{J}(r)=\mbf{D}(r)
\mbf{M}(r)$. Note that $\mbf{J}(r)$ is always upper-triangular. This follows
from the upper-triangular structure of $\mbf{D}(r)$ and of the lower-triangular
structure of $\mbf{M}(r)$. Next we rewrite the vector $\mbf{\beta}(r,n)$. We
use an identity from the calculus of exterior derivatives for
pseudo-differential operators \cite{manin1}:
\beq\la{manin1}
d \beta_{-1}(r+n)=\frac{r+n}{r} \sum_{j=-1-n}^{r-2} \beta_{-1-j}(n) d
\alpha_{j}(r),
\eeq

\no which gives
\beq\la{manin2}
\beta_{-j}(n)=\frac{r}{r+n}\dd{\beta_{-1}(r+n)}{\alpha_{j-1}(r)},
\eeq

\no where $\delta/\delta \alpha_{j-1}(r)$ is the {\em variational derivative}
with respect to $\alpha_{j-1}(r)$. Hence \beq\la{maninfinal}
\mbf{\beta}(r,n)=\frac{r}{r+n} \dd{}{\mbf{\alpha}(r)} \beta_{-1}(r+n). \eeq

\no Equations \rf{first} become
\beq\la{second}
\pp{\mbf{\alpha}(r)}{t_n}=\frac{r}{r+n}\mbf{J}(r) 
\dd{}{\mbf{\alpha}(r)} \beta_{-1}(r+n).
\eeq

\no This set of equations is Hamiltonian \cite{strampp}, with Hamiltonian 
\beq\la{hamil}
H(r,n)=\frac{r}{r+n} \beta_{-1}(r+n)=\frac{r}{r+n} \alpha_{-1}(r+n).
\eeq

\no (We have used the observation that $\alpha_{-1}(r)=\beta_{-1}(r)$.) It
suffices to prove that the operator $J(r)$ is Hamiltonian \cite{anton}, 
$i.e.$, that the operator $\mbf{J}(r)$ defines a Poisson bracket. This Poisson
bracket is given by \cite{strampp}
\beq\la{pbpde}
\{S,T\}=\left(\dd{S}{\mbf{\alpha}(r)}\right)^T \mbf{J}(r)
\left(\dd{T}{\mbf{\alpha}(r)}\right).
\eeq

Denote by ${\cal H}$ the quotient space of all smooth functionals of the
potentials in $\mbf{\alpha}(r)$, modulo total derivatives with respect to $x$.
\no For \rf{pbpde} to define a Poisson bracket on ${\cal H}\times {\cal H}$, we
need three properties:

\begin{enumerate}

\item {\bf bilinearity:} This is obvious.

\item {\bf skew-symmetry:} This is less obvious. Notice that the functions
appearing in the bracket \rf{pbpde} appear only through their variational
derivatives. Hence, these functions are only defined up to total derivatives
with respect to $x$, $i.e.$, they are elements of $\cal H$.  The
skew-symmetry of the Poisson bracket \rf{pbpde} operating on ${\cal H}\times {\cal
H} $ is then easily obtained by integration by parts.

\item {\bf Jacobi identity:} This is also not obvious. The proof can be found
in \cite{strampp}. There it is shown that the above bracket is essentially the
bracket Adler defines in \cite{adler}. The proof of the Jacobi identity then
reduces to Adler's proof. 

\end{enumerate}

The Hamiltonian system of PDE's \rf{second} describes a whole hierarchy of
Hamiltonian PDEs for a fixed $r$. All the members of the hierarchy have the
same Poisson structure with different Hamiltonians: for each $n$ coprime with
$r$, a different system of Hamiltonian partial differential equations is
obtained, describing the $t_n$-evolution of the potentials. Note that the
first member of every hierarchy is trivial. From the first flow of \rf{lax},
we get
\beq\la{trivial}
\pp{L}{t_1}=\left[L_+,L\right]=\left[\p,L\right]=\p L-L \p=\pp{L}{x}+L \p-L
\p=\pp{L}{x}. 
\eeq

\no which is the same for all $r$-reductions. Hence, the first member of every
hierarchy is $\p \mbf{\alpha}(r)/\p t_1=\p \mbf{\alpha}(r)/\p x$.

For example, choosing $r=2$ results in the KdV hierarchy with Poisson operator
$J(2)=2 \p$. Choosing $r=3$ gives the Boussinesq hierarchy \cite{zak,mckean1} with
Poisson operator

\beq\la{operjbous}
J(3)=\left(\ba{cc}0 & 3 \p\\3 \p & 0\ea\right).
\eeq

Some remarks are in order about the Hamiltonian PDE's
\beq\la{hamsys}
\pp{\mbf{\alpha}(r)}{t_n}=\mbf{J}(r) \dd{H(r,n)}{\mbf{\alpha}(r)}.
\eeq

\no {\bf Remarks}

\begin{description}

\item[(a)~] Note that in contrast with the Poisson brackets in \cite{gardner,
zak, mckean1}, the bracket \rf{pbpde} is local, $i.e.$, it does not involve
integrations. This  is an immediate consequence of working in the quotient
space ${\cal H}$.  The Poisson bracket \rf{pbpde} for $r=2$ and $r=3$ are the
integrands of the  brackets introduced in \cite{gardner} and \cite{zak,
mckean1} respectively. 

\item[(b)~] Bogoyavlenskii and Novikov \cite{bogoyavlenskii} considered only the
Korteweg-deVries equation. As a consequence, none of the algebraic machinery of
this and the previous section was required for their approach. Their starting point was
the KdV equivalent of \rf{hamsys}. This same starting point is used if one
considers any other integrable partial differential equation which has only one
spatial dimension, such as the nonlinear Schr\"{o}dinger equation, the modified
Korteweg-deVries equation, etc. By starting with a one-dimensional partial
differential equation, the first step (imposing the $r$-reduction) is skipped. It
is for this step that the algebraic language of the previous two sections is
required. 

\item[(c)~] An infinite number of conserved quantities exists for each member of the
hierarchy given by \rf{hamsys}. This is a necessary condition for the
integrability of these partial differential equations.  Adler \cite{adler}
showed that the different members of the $r$-reduced hierarchy define mutually
commuting flows. The infinite set of Hamiltonians $\{H(r,n): n\geq
1\}$ is a set of conserved densities for every member of the hierarchy. That
different members of the hierarchy \rf{hamsys} commute is expressed as the
commutation of their respective Hamiltonians under the Poisson bracket \rf{pbpde}:
\beq\la{pbcommute}
\left\{H(r,k_1),H(r,k_2)\right\}=0,
\eeq
for a fixed $r$ and all $k_1, k_2$.

Denote the solution operator of the $n$-th member of the hierarchy \rf{hamsys}
by $\mbf{K}_n(t_n)$. In other words, given initial conditions 
$\mbf{\alpha}(r)(x,t_n=0)$,
the solution $\mbf{\alpha}(r)(x,t_n)$ for any $t_n$ is written as
\beq
\mbf{\alpha}(r)(x,t_n)=\mbf{K}_n(t_n) \mbf{\alpha}(r)(x,0).
\eeq

\no Adler's statement \cite{adler} that different flows in the hierarchy
\rf{hamsys} commute is then expressed as
\beq
\mbf{K}_n(t_n) \mbf{K}_m(t_m)=\mbf{K}_m(t_m) \mbf{K}_n(t_n).
\eeq

\item[(d)~] The Hamiltonian operator $\mbf{J}(r)$ is usually degenerate,
$i.e.$, its kernel is not empty. Adler \cite{adler} showed that the
variational derivatives of the elements of the set \linebreak
$\{\alpha_{-1}(r+n): -r+1 \leq  n\leq -1\}$ are all annihilated by
$\mbf{J}(r)$. In other words, these elements are Casimir functionals for the
flows generated by the $r$-reduction. It is easy to see from the triangular
form of $\mbf{J}(r)$ that the dimension of its kernel is exactly $r-1$. This
implies that the set of Casimir functionals found by Adler forms a complete
basis for the kernel of $\mbf{J}(r)$ (see also \cite{veselov}).

\end{description}




\section{Impose the $n$-reduction}\la{sec:nred}

Next, we consider stationary solutions of the system \rf{hamsys}, for the $n$
value determined in step 2. Hence, from \rf{stat},

\beq\la{stathamsys}
\sum_{k=1}^n d_k \mbf{J}(r) \dd{H(r,k)}{\mbf{\alpha}(r)}=0
~\Rightarrow~\mbf{J}(r) \dd{}{\mbf{\alpha}(r)}\sum_{k=1}^n d_k H(r,k)=0,
\eeq

\no with $d_n=1$. Furthermore, without loss of generality, $d_{n-1}$ can be
equated to 0, as was shown in \cite{krich4, ds1}. 

The following theorem was first proved for the KdV equation by Lax \cite{lax1}
and Novikov \cite{novikov}. 

\begin{theo}\la{theo:statpoint}
The set of stationary solutions with respect to $t_n$ is invariant under the
action of any of the other higher-order flows. 
\end{theo}

\no {\bf Proof} Consider the hierarchy of mutually commuting
Hamiltonian systems 
\beq\la{hamsysgen}
\pp{\mbf{\alpha}(r)}{\tilde{t}_n}=\mbf{J}(r)\dd{}{\mbf{\alpha}(r)}\left(\sum_{k=1}^{n}
d_k H(r,k)\right).
\eeq
Denote the solution operator of the $n$-th member of this hierarchy as
$\tilde{\mbf{K}}_n(\tilde{t}_n)$. Clearly these solution operators commute with
the solution operators of the higher-order flows, $\mbf{K}_m(t_m)$:
\beq\la{comgen}
\tilde{\mbf{K}}_n(\tilde{t}_n) \mbf{K}_m(t_m)=
\mbf{K}_m(t_m) \tilde{\mbf{K}}_n(\tilde{t}_n).
\eeq
\no A stationary solution with respect to $t_n$ is a fixed point of the
operator $\tilde{\mbf{K}}_n(\tilde{t}_n)$:\linebreak
$\tilde{\mbf{K}}_n(\tilde{t}_n) \mbf{\alpha}(r)(x)=\mbf{\alpha}(r)(x)$. Hence
\bea
\mbf{K}_m(t_m)\tilde{\mbf{K}}_n(\tilde{t}_n) \mbf{\alpha}(r)(x)&=&
\mbf{K}_m(t_m)\mbf{\alpha}(r)(x)\\
\Rightarrow~~\tilde{\mbf{K}}_n(\tilde{t}_n)\mbf{K}_m(t_m) \mbf{\alpha}(r)(x)&=&
\mbf{K}_m(t_m)\mbf{\alpha}(r)(x),
\eea
\no since the two operators commute. Hence $\mbf{K}_m(t_m)\mbf{\alpha}(r)(x)$ is
a fixed point of $\tilde{\mbf{K}}_n(\tilde{t}_n)$ and hence a stationary
solution with respect to $t_n$. \hspace*{\fill}$\bbox$

Let us examine the structure of the equations \rf{stathamsys}, determining the
stationary solutions with respect to $t_n$. From \rf{stathamsys}, 
$\dd{}{\mbf{\alpha}(r)}\sum_{k=1}^n d_k H(r,k)$ is in the kernel of the Poisson
operator $\mbf{J}(r)$. Hence it is a linear combination of the Casimir
functionals:

\beq\la{cashamsys}
\dd{}{\mbf{\alpha}(r)}\sum_{k=1}^n d_k H(r,k)+\sum_{k=1}^{r-1}h_k \frac{r}{k}
\dd{\alpha_{-1}(k)}{\mbf{\alpha}(r)}=0;
\eeq

\no the coefficient of the Casimir functional $\alpha_{-1}(r)$ has been
written as $h_k r/k$ for convenience. Equation \rf{cashamsys} is a system of
Euler-Lagrange equations, with Lagrangian depending on the $r-1$ potentials
$\mbf{\alpha}(r)$ and their derivatives:

\bea\nonumber
{\cal L}(r,n)&=&H(r,n)+\sum_{k=1}^{n-2}d_k H(r,k)+\sum_{k=1}^{r-1} h_k \frac{r}{k}
\alpha_{-1}(k)\\\la{lagrangian}
&=&H(r,n)+\sum_{k=1}^{n-2}d_k H(r,k)+\sum_{k=1}^{r-1} h_k H(r,k-r),
\eea

\no since $d_{n-1}\equiv 0$. The last term in this equation is a slight abuse
of notation. It is to be interpreted using \rf{hamil}. 

The set of $r-1$ Euler-Lagrange equations \rf{cashamsys}

\beq\la{el}
\dd{{\cal L}(r,n)}{\mbf{\alpha}(r)}=0
\eeq

\no will be referred to as the {\em $(r,n)$-th (stationary) KP equation}. This
system of Euler-Lagrange equations is extremely important: it is a
finite-dimensional system of {\em ordinary differential equations} describing
how solutions of the $(r,n)$-th KP equation depend on $x$. At this point, the
study of rank 1, finite-genus solutions of (KP) is immensely simplified: it is
reduced to the study of a finite-dimensional set of ordinary differential
equations \rf{el} that are derived from one scalar quantity, the
Lagrangian \rf{lagrangian}. The remainder of this paper examines the special
structure of the set of equations \rf{el}. 

\vspace*{12pt}
\no {\bf Remarks}

\begin{description}

\item[(a)~] As pointed out before, the first flow of the KP hierarchy defines
$t_1$ to be $x$. This first flow imposes no constraints on the $x$-dependence
of the potentials. After one imposes the $r$- and $n$-reductions this
$x$-dependence is determined by the Lagrangian system \rf{el}. 

\item[(b)~] The Euler-Lagrange equations are a {\em minimal} set of
differential equations the potentials in $\mbf{\alpha}(r)$ have to satisfy to
make the $t_r$-flow and the $t_n$-flow stationary. In step 5 of \cite{ds1}, a
system of differential equations was proposed which the initial conditions of
the KP equation needs to satisfy. Because of the way the $r$- and
$n$-reductions were performed, it was not clear in \cite{ds1} that these
differential equations are in fact {\em ordinary} differential equations.
Furthermore, the differential equations obtained in \cite{ds1} were not
necessarily functionally independent. The equations \rf{el} are functionally
independent, as was shown by Veselov \cite{veselov}. 

\item[(c)~]
Since the order of imposing the $r$-reduction and the $n$-reduction can be
reversed (in \cite{ds1} they were executed simultaneously), the remark on
page~\pageref{p:rem2}
can be repeated with $n$ instead of $r$: if we impose an $n$-reduction, we
automatically achieve an $nk$ reduction for any positive integer $k$. Imposing
an $n$-reduction implies that $L^n$ is purely differential. In that case
$L^{nk}=(L^n)^k$ is also purely differential.

\item[(d)~] At this point, {\em Krichever's criterion \cite{krichcom}}
surfaces\footnote{There is some discussion in the literature about the
necessity of this criterion. See for instance \cite{kasman}}: for a
finite-genus solution of rank 1 of the KP equation, $r$ and $n$ need to be
coprime. Non-coprime $r$ and $n$ result in higher-rank solutions.  We show
next that to determine a rank 1, finite-genus solution completely, non-coprime
$r$ and $n$ are not allowed. 

Imposing the $r$- and $n$-reductions amounts (up to including lower order flows)
to imposing that both $L^r$ and $L^n$ are purely differential operators. If $r$
and $n$ are not coprime, let $r=k\hat{r}$ and $n=k \hat{n}$, for integers $k,
\hat{r}$ and $\hat{n}$. So, $r$ and $n$ have the common factor $k$. If
$L^k=L^k_+$ ($i.e.$, $L^k$ is purely differential) then the solution is
stationary with respect to $t_k$. Since $L^r=(L^k)^{\hat{r}}$ and
$L^n=(L^k)^{\hat{n}}$ are purely differential, the solution is trivially
stationary with respect to $t_r$ and $t_n$. Thus, imposing stationarity with
respect to $t_k$ implies stationarity with respect to $t_r$ and $t_n$. Imposing
stationarity with respect to only one higher order flow $t_k$ however, does not
provide enough information for the determination of the solution using our
methods. Therefore, $r$ and $n$ are required to be coprime. 

\end{description}

\section{The explicit dependence of the Lagrangian on the potentials and their
derivatives}\la{sec:exp}

We want to examine the explicit functional dependence of the Lagrangian ${\cal
L}(r,n)$ on the potentials $\mbf{\alpha}(r)=(\alpha_0(r), \alpha_1(r), \ldots,
\alpha_{r-2}(r))^T$ and their derivatives. We are especially interested in the
order of the highest derivative of $\alpha_i(r)$, for $i=0, 1, \ldots, r-2$.
This information is necessary in order to carry out the generalization of the 
Legendre transformation in Section \ref{sec:ostro}.

\vspace*{12pt}
\no {\bf Definition:}
{\bf The weight of an operator f(x), W[f(x)]}, is defined to be an
integer-valued functional with the following properties:

\been



\item $W[f g]=W[f]+W[g]$, and $W[f^N]=N W[f]$ for integer N.

\item If $W[f]=W[g]$ then $W[f \pm g]=W[f]=W[g]$. 

\item $W[L]=1$.

\een
\vspace*{12pt}

The usefulness of this definition is connected with the scaling symmetry of
the KP equation. This symmetry is shared by the whole KP hierarchy through its
definition using the operator $L$. Introducing the weight function turns the
algebra of operators used here into a so-called graded algebra \cite{dickey}.
Essentially, the weight function introduced here is identical with the `rank'
introduced by Miura, Gardner and Kruskal in \cite{mgk}. We use the name
`weight' because `rank' has a very different meaning in KP theory \cite{nmpz,
krichcom}.

Using the defining properties of the weight function, we calculate the weight
of some quantities we have used:

\vspace*{12pt} 
\no {\bf Examples}
\begin{itemize}

\item Since $W[L]=1$, any term in $L$ has weight 1. In particular $W[\p]=1$.

\item Hence $W[u_k \p^{-k+1}]$ $=$ $1 \Rightarrow W[u_k]+W[\p^{-k+1}]=1
\Rightarrow$ \newline
$W[u_k]+(-k+1)W[\p]=1 \Rightarrow W[u_k]=k$.

\item $W[L^r]=r$, hence $W[\alpha_k(r) \p^k]=r~\Rightarrow$~
$W[\alpha_k(r)]+W[\p^k]=r~\Rightarrow~W[\alpha_k(r)]=r-k$. Analogously
$W[\beta_{k}(r)]=r-k$. 

\item $W[\p/\p t_r]=W[L^r]=r$.

\item $W[H(r,n)]=r+n+1$, from \rf{hamil}. Then also, $W[{\cal L}(r,n)]=r+n+1$. 

\item $W[d_k]=n-k$ and $W[h_k]=r+n-k$.

\end{itemize}

\vspace*{12pt}

We now use the weight function to calculate how the Lagrangian depends on the
$r-1$ potentials in $\alpha(r)$ and their derivatives.  For $j=2, 3, \ldots,
r$, let us denote by $N_j$ the highest order of differentiation of the
potential $\alpha_{r-j}(r)$ in the Lagrangian \rf{lagrangian}. We say a term
in the Lagrangian is of degree $M$ if it contains $M$ factors (counting
multiplicities) that are linear in one of the potentials or in one of their
derivatives. The Lagrangian has linear terms ($i.e.$, terms of degree one),
quadratic terms ($i.e.$, terms of degree two), cubic terms ( $i.e.$, terms of
degree three), terms of degree four and so on. Clearly the linear terms can
not depend on the derivatives of the potentials: such a term would be a total
derivative and would be disregarded.

All terms in the Lagrangian have the same weight, as a consequence of the
scaling invariance of the KP hierarchy. In order to find the highest derivative
of a potential, we need only consider the quadratic terms of the Lagrangians.
All higher degree terms have contributions from more than one other potential.
The nontrivial potential with the lowest weight is $\alpha_{r-2}(r)$:
$W[\alpha_{r-2}(r)]=2$. Every time a potential
appears in a term of the Lagrangian, the number of $x$-derivatives of the other
potentials in that term decreases by at least 2. Since the linear terms cannot
contain derivatives, it follows that the highest derivatives of the potentials
are found in the quadratic terms of the Lagrangian.

Similarly, every time one of the coefficients $h_k$, for $k=1, 2,
\ldots, r-1$ or $d_j$ for $j=1,2,\ldots, n-2$ appears in a term of the
Lagrangian, the number of $x$-derivatives in that term has to decrease by the
weight of the coefficient $h_k$ or $d_j$ involved. It follows that the highest
derivatives of the potentials are found in the quadratic terms of the
Lagrangian, not containing any of these coefficients, $i.e.$, in the
quadratic terms of $H(r,n)$. 

We use these observations to find how the Lagrangian depends on the highest
derivatives of the potentials, $i.e.$, to find $N_j$, for $j=2,3,\ldots,
r$.

\begin{theo}\la{theo:lagdep}
The order of differentiation in the Lagrangian ${\cal L}(r,n)$ of
$\alpha_{r-2}(r)$ is $[(r+n-3)/2]$, $i.e.$, $N_2=[(r+n-3)/2]$. The order of
differentiation in the Lagrangian ${\cal L}(r,n)$ of any other potential
$\alpha_{r-i}(r)$ is $[(r+n-2i+2)/2]$, $i.e.$, $N_i=[(r+n-2i+2)/2]$ for
$i=3,4,\ldots, r$. The square brackets denote the integer part of the
expression inside.
\end{theo}

\no {\bf Proof} 

The proof is a tedious check on all the possibilities of
combinations of derivatives of the potentials appearing in the Lagrangian
${\cal L}(r,n)$. We start with a few specific cases before examining the
general case. 

\vspace*{0.5cm}

\no{\em Dependence of ${\cal L}(r,n)$ on $\alpha_{r-2}(r)$}\vspace{0.5cm}

We consider terms of the form $\aa{r-2}{k_1} \aa{r-j}{k_2}$. We want to
find the maximal allowable value for $k_1$, $i.e.$, for the highest
derivative of $\alpha_{r-2}(r)$ appearing in the Lagrangian. We have
\beqno
W[\aa{r-2}{k_1} \aa{r-j}{k_2}]=k_1+k_2+2+j=r+n+1=W[{\cal L}(r,n)], 
\eeqno

\no hence 
\beqno
k_1+k_2=r+n-1-j.
\eeqno

Only values of $k_1, k_2$ with $|k_1-k_2|\leq 1$ need to be considered in this
case. Other values are reduced to these cases using integration by parts. If
$j=2$, then necessarily $k_1=k_2$. Otherwise the term we are considering is a
total derivative, equivalent to $([\aa{r-2}{k_2}]^2/2)'$. In this case we find
\beqno
k_1=\frac{r+n-3}{2}.
\eeqno

\no This value is not necessarily an integer. If it is, when $r+n-3$ is even,
it would be the maximum value for $k_1$. Otherwise, this term does not appear
in ${\cal L}(r,n)$, so we consider next $j=3$. If $k_1=k_2+1$, we find
$k_1=(r+n-5)/2$. If this is an integer, than so is $(r+n-3)/2$, hence this does
not raise the order of differentiation with which $\alpha_{r-2}(r)$ appears. On
the other hand, if $k_1=k_2$, we find
\beqno
k_1=\frac{r+n-4}{2}. 
\eeqno

\no Either $(r+n-3)/2$ or $(r+n-4)/2$ is guaranteed to be an integer. This
integer  is the maximal order of differentiation with which $\alpha_{r-2}(r)$
appears in the Lagrangian: hence
\beq\la{N2}
N_2=\left[\frac{r+n-3}{2}\right],
\eeq

\no where the square brackets denote the integer part of the
expression inside. If $r+n-3$ is even, this results in the first value we
obtained for $k_1$. If $r+n-3$ is odd, we find the second expression. 

\vspace*{0.5cm}

\no{\em Dependence of ${\cal L}(r,n)$ on $\alpha_{r-3}(r)$}\vspace{0.5cm}

Next consider terms of the form $\aa{r-3}{k_1} \aa{r-j}{k_2}$. We want to
find the maximal allowable value for $k_1$, $i.e.$, $N_3$. We have
\beqno
W[\aa{r-3}{k_1} \aa{r-j}{k_2}]=k_1+k_2+3+j=r+n+1=W[{\cal L}(r,n)], 
\eeqno

\no or 
\beqno
k_1+k_2=r+n-2-j.
\eeqno

If $j=2$, then for the case $k_1=k_2$, we find 
\beqno
k_1=\frac{r+n-4}{2}.
\eeqno

\no In the other case, $k_2=k_1+1$ (we can always write the Lagrangian such
that the potentials with the lowest weight have the higher order of
differentiation), we obtain
\beqno
k_1=\frac{r+n-5}{2}.
\eeqno 

\no In this case, $k_2=(r+n-3)/2$, which corresponds to $N_2$ (if $r+n-3$ is
even), found in the previous section.

The analysis for $j>2$ does not increase the possible values of $k_1$. Either
$(r+n-4)$ or $(r+n-5)$ is guaranteed to be even, so
\beq\la{N3}
N_3=\left[\frac{r+n-4}{2}\right].
\eeq

\vspace*{0.5cm}

\no{\em Dependence of ${\cal L}(r,n)$ on $\alpha_{r-4}(r)$}\vspace{0.5cm}

Consider terms of the form $\aa{r-4}{k_1} \aa{r-j}{k_2}$. We want to
find the maximal allowable value for $k_1$, $i.e.$, $N_4$. We have
\beqno
W[\aa{r-4}{k_1} \aa{r-j}{k_2}]=k_1+k_2+4+j=r+n+1=W[{\cal L}(r,n)], 
\eeqno

\no or 
\beqno
k_1+k_2=r+n-3-j.
\eeqno

Consider the case when $j=2$. If $k_1=k_2$, then $k_1=(r+n-5)/2$ and
$k_2=(r+n-5)/2=(r+n-3)/2-1$. If $r+n-5$ is an integer, then so is $r+n-3$. In
this case we can use integration by parts to decrease $k_1$ by 1 and increase
$k_2$ by 1. Therefore this possibility is not allowed. If $k_2=k_1+1$, then we
obtain 
\beqno
k_1=\frac{r+n-6}{2} ~~\mbox{and}~~ k_2=\frac{r+n-4}{2}. 
\eeqno

\no This possibility is allowed. Also, if we let $k_2=k_1+1$, we find 
\beqno
k_1=\frac{r+n-7}{2} ~~\mbox{and}~~ k_2=\frac{r+n-3}{2}. 
\eeqno

\no This possibility is also allowed. Examining the other possibilities for $j,
k_2$ does not improve the allowed values for $k_1$, therefore
\beq\la{N4}
N_4=\left[\frac{r+n-6}{2}\right].
\eeq

\no{\em Dependence of ${\cal L}(r,n)$ on $\alpha_{r-i}(r)$, $3\leq i
\leq r$}\vspace{0.5cm}

We now state the general case. Using arguments as above (we want potentials
with lower weight to have a higher order of differentiation in each term of the
Lagrangian), we have 
\beqno
W[\aa{r-i}{k_1} \aa{r-j}{k_2}]=k_1+k_2+i+j=r+n+1=W[{\cal L}(r,n)], 
\eeqno

\no or 
\beqno
k_1+k_2=r+n+1-i-j.
\eeqno

Consider the case when $j=2$. Then if $k_2=k_1+m$, 
\beqno
k_1=\frac{r+n-1-i-m}{2} ~~\mbox{and}~~ k_2=\frac{r+n-1-i+m}{2}. 
\eeqno

\no Using the above argument, we obtain an allowed possibility if either
$k_2=(n+r-3)/2$ or $k_2=(n+r-4)/2$. This gives two possible values for $m$:
\beqno
m=i-2 ~~ \mbox{or} ~~ m=i-3.
\eeqno

\no These respectively give
\beqno
k_1=\frac{r+n-2i+1}{2} ~~\mbox{and}~~ k_2=\frac{r+n-2i+2}{2}.
\eeqno

The other possibilities for $j$ give no additional information, hence
\beq\la{Ni}
N_i=\left[\frac{n+r-2i+2}{2}\right].
\eeq

This formula is valid for all $i$, except $i=2$, the first value.
\hspace*{\fill}$\bbox$ 

Table \ref{table1} gives an overview of the possibilities. 

\settowidth{\mylength}{$\left[\frac{r+n-3}{2}\right]$}
\settoheight{\myheight}{$\left[\frac{r+n-3}{2}\right]$}
\addtolength{\myheight}{8pt}

\begin{table}[htb]
\begin{center}
\caption{\bf The order of differentiation with which the potentials appear 
in the Lagrangian \la{table1}}
\vspace*{0.2in}
\begin{tabular}{|c|c|}
\hline
$i$ & $N_i$ \\
\hline\hline
\pb{2} & $\left[\frac{r+n-3}{2}\right]$ \\
\hline
\pb{3} & $\left[\frac{r+n-4}{2}\right]$ \\
\hline
\pb{$\vdots$} & $\vdots$ \\
\hline
\pb{$j$} & $\left[\frac{r+n-2 j+2}{2}\right]$ \\
\hline
\pb{$\vdots$} & $\vdots$ \\
\hline
\pb{$r$} & $\left[\frac{n-r+2}{2}\right]$\\
\hline
\end{tabular}
\end{center}
\end{table}

\vspace*{12pt}
\no{\bf Remark}
In \cite{ds1}, it
was argued that a generic rank 1, genus $g$ solution of the KP equation
corresponds to a solution of the $(r,n)$-th KP equation with $r=g+1$ and
$n=g+2$. The dependence of the Lagrangian ${\cal L}(r,n)={\cal L}(g+1,g+2)$ on
the potentials for this case is found from Table \ref{table1} by using these
values for $r$ and $n$. It follows that in the generic case, the potential
$\alpha_{r-j}(r)$ appears with $g-j+2$ derivatives, for $j=2,3,\ldots, g+1$. 

\section{The Ostrogradskii transformation, canonical variables and the
Hamiltonian system}\la{sec:ostro}

If we have a Lagrangian system, where the Lagrangian only depends on the
potentials and their first derivatives, then under certain conditions we can
use the Legendre transformation \cite{arnold} to write the Lagrangian system in
first-order form as a Hamiltonian system with canonical variables
$(\mbf{q},\mbf{p})$. Here the variables $\mbf{q}$ are the potentials appearing
in the Lagrangian and all of their derivatives, except the highest derivative.
Their {\em conjugate variables} $\mbf{p}$ are defined as the partial
derivatives of the Lagrangian with respect to the $\mbf{q}$ variables
\cite{arnold}.

The Lagrangian system \rf{el} we constructed from the KP hierarchy, assuming
two of its flows are stationary, depends on more than just the first
derivatives of the potentials $\mbf{\alpha}(r) = (\alpha_{0}(r), \alpha_1(r),
\ldots, $ $\alpha_{r-2}(r))^T$. The Legendre transformation is generalized to
write the Lagrangian system \rf{el} in Hamiltonian form. This is achieved by
Ostrogradskii's theorem, given later in this section. Consider the
Ostrogradskii transformation (see \cite{dfn2, whittaker} for a simpler
version) \beq\la{ostrotrans}  q_{ij}=\ppn{j-1}{}{x}\alpha_{r-i-1}(r),
~~p_{ij}=\dd{{\cal L}(r,n)}{\aa{r-i-1}{j}} \eeq

\no for $i=1,2, \ldots, r-1$ and $j=1, 2, \ldots, N_{i+1}$.

Note that when all the $N_j=1$, for $j=2,3,\ldots, r$ ($i.e.$, when the
Lagrangian only depends on first derivatives), then the Ostrogradskii
transformation \rf{ostrotrans} reduces to the Legendre transformation.

Using the definition of the variational derivative, we establish the recurrence
relations 
\beq\la{ostrorecur}
p_{ij}=\pp{{\cal L}(r,n)}{\aa{r-i-1}{j}}-\pp{p_{i(j+1)}}{x}, ~~\mbox{for} ~~
j=1,2,\ldots, N_{i+1}.
\eeq

Here we have defined $p_{i(N_{i+1}+1)}$ to be zero. These recurrence relations
will be useful later on. Furthermore, from the definition of the Ostrogradskii
transformation \rf{ostrotrans}, we obtain
\beq\la{ostroweight}
W[q_{ij}]=i+j~~\mbox{and}~~W[p_{ij}]=r+n-(i+j).
\eeq

\no Weight relationships such as these provide a quick check for the validity of
large expressions, such as the ones encountered below. Many typographical
errors are easily avoided by checking that only terms with the same weight are
added.

The Lagrangian needs to fulfill a nonsingularity requirement for the
Ostrogradskii transformation to be invertible:

\vspace*{12pt}
\no {\bf Definition \cite{dfn2}:}
The Lagrangian ${\cal L}(r,n)$ is {\bf (strongly) nonsingular} if
the Ostrogradskii transformation \rf{ostrotrans} can be solved in the form
\beqno
\aa{r-i}{j}=\aa{r-i}{j}(\mbf{q},\mbf{p}), ~~\mbox{for}~i=2,3,\ldots, r
~~\mbox{and}~j=1,2,\ldots, 2 N_i-1. 
\eeqno

\no Here $\mbf{q}$ and $\mbf{p}$ denote the vector of all variables $q_{ij}$
and $p_{ij}$ respectively. In other words, the Ostrogradskii transformation
\rf{ostrotrans} is invertible if and only if the Lagrangian is nonsingular.
Then the Euler-Lagrange equation \rf{el} can be written in first-order
form using the variables $\mbf{q}$ and $\mbf{p}$.

Define the vector $\mbf{X}=(\aa{r-2}{N_2}, \aa{r-3}{N_3}, \ldots,
\aa{0}{N_r})^T$. This is the vector containing the highest derivatives of the
potentials.  We already know that the highest derivatives of the potentials
are found in the quadratic terms of the Lagrangian. The Lagrangian is
conveniently written in the following form:
\beq\la{quadraticform}
{\cal L}(r,n)=\frac{1}{2} \mbf{X}^T \mbf{{\cal G}}(r,n) \mbf{X}+\mbf{{\cal
A}}^T (r,n) \mbf{X}+  \tilde{\mbf{{\cal L}}}(r,n),
\eeq

\no with $\mbf{{\cal G}}(r,n)$, $\mbf{{\cal A}}(r,n)$, $\tilde{{\cal L}}(r,n)$
independent of $\mbf{X}$. $\mbf{{\cal G}}(r,n)$ is a constant symmetric $(r-1)
\times (r-1)$ matrix. $\mbf{{\cal A}}(r,n)$ is an $(r-1)$-dimensional
vector.In the classical case of the Legendre transformation $\mbf{{\cal
G}}(r,n)$ can be regarded as either a metric tensor or as the inverse of a
mass tensor \cite{arnold}.

The following theorem generalizes a well-known result for the Legendre
transformation.

\begin{theo}\la{prop:sing}
The Lagrangian ${\cal L}(r,n)$ is nonsingular if and only if 
the matrix $\mbf{{\cal G}}(r,n)$ in \rf{quadraticform} is nonsingular. 
\end{theo}

\no{\bf Proof} 
The proof is an extension of the proof in the case when the Lagrangian depends
on only one potential \cite{dfn2}. We demonstrate that under the assumption
that $\mbf{{\cal G}}(r,n)$ is nonsingular, the Ostrogradskii transformation
\rf{ostrotrans} is invertible. Then by definition the Lagrangian
is nonsingular. 

First note that the variables $\mbf{q}$ are expressed in terms of the
potential and its derivatives, by their definition \rf{ostrotrans}.
Furthermore, the Lagrangian ${\cal L}(r,n)$ is a function of only $\mbf{q}$
and  $\mbf{X}$: it follows from the Ostrogradksii transformation
\rf{ostrotrans} that all derivatives of the potentials in the Lagrangian are
$\mbf{q}$-variables, except the highest derivative of the potentials. These are
the components of $\mbf{X}$.

We now construct the vector 
\begin{eqnarray*}
\mbf{P}&=&\left(
\ba{cccc}
p_{1N_2},&
p_{2N_3},&
\cdots,&
p_{(r-1)N_r}
\ea
\right)^T\\
&=&
\left(
\dd{{\cal L}(r,n)}{\aa{r-2}{N_2}},
\dd{{\cal L}(r,n)}{\aa{r-3}{N_3}},
\cdots,
\dd{{\cal L}(r,n)}{\aa{0}{N_r}}
\right)^T\\&=&
\dd{{\cal L}(r,n)}{\mbf{X}}.
\end{eqnarray*}

\no Since $\mbf{X}$ contains the highest derivatives of the potentials, by 
definition of the variational derivative we get
\begin{eqnarray}\la{labeliguess}
\left(
\ba{c}
p_{1N_2}\\
p_{2N_3}\\
\vdots\\
p_{(r-1)N_r}
\ea
\right)=\pp{{\cal L}(r,n)}{\mbf{X}}=\mbf{{\cal G}}(r,n) \mbf{X}+
\mbf{{\cal A}}(r,n). 
\end{eqnarray}

\no Now we solve \rf{labeliguess} for $\mbf{X}$, since by assumption
$\mbf{{\cal G}}(r,n)$ is nonsingular. We denote $\mbf{{\cal
G}}^{-1}(r,n)=\mbf{{\cal M}}(r,n)$. Since  $\mbf{{\cal G}}(r,n)$ is symmetric,
so is $\mbf{{\cal M}}(r,n)$.
\bea\nonumber
\mbf{X}&=&\left(
\ba{c}
\aa{r-2}{N_2}\\
\aa{r-3}{N_3}\\
\vdots\\
\aa{0}{N_r}
\ea
\right)=\mbf{\cal M}(r,n)
\left(
\ba{c}
p_{1N_2}\\
p_{2N_3}\\
\vdots\\
p_{(r-1)N_r}
\ea
\right)-\mbf{\cal M}(r,n)\mbf{\cal A}(r,n)\\\la{xintermsofp}
&=&\mbf{\cal M}(r,n)\left(\mbf{P}-\mbf{\cal A}(r,n)\right).
\eea

We have expressed $\mbf{X}$ in terms of the coordinates $\mbf{q}$ and
$\mbf{p}$. We want to do the same with its derivatives. Now consider the
following set of Ostrogradskii equations, using the recurrence relations
\rf{ostrorecur}
\begin{eqnarray*} \left( \ba{c} p_{1(N_2-1)}\\ p_{2(N_3-1)}\\ \vdots\\
p_{(r-1)(N_r-1)} \ea \right)&=& \left( \ba{c} \pp{{\cal
L}(r,n)}{\aa{r-2}{N_2-1}}-\pp{p_{1N_2}}{x}\\ \pp{{\cal
L}(r,n)}{\aa{r-3}{N_3-1}}-\pp{p_{2N_3}}{x}\\ \vdots\\ \pp{{\cal
L}(r,n)}{\aa{0}{N_r-1}}-\pp{p_{(r-1)N_r}}{x} \ea \right)\\&=& \left( \ba{c}
\pp{{\cal L}(r,n)}{\aa{r-2}{N_2-1}}\\ \pp{{\cal L}(r,n)}{\aa{r-3}{N_3-1}}\\
\vdots\\ \pp{{\cal L}(r,n)}{\aa{0}{N_r-1}} \ea \right)-\pp{}{x} \left( \ba{c}
p_{1N_2}\\ p_{2N_3}\\ \vdots\\ p_{(r-1)N_r} \ea \right). \end{eqnarray*}

\no Note that the first term depends only on $\mbf{q}$ and $\mbf{X}$.  The last
term is the derivative of $\mbf{\cal G}(r,n) \mbf{X}+\mbf{\cal
A}(r,n)$. Since $\mbf{\cal A}(r,n)$ depends only on
$\mbf{q}$, the derivative of $\mbf{\cal A}(r,n)$ depends only on
$\mbf{q}$ and on $\mbf{X}$.  Since $\mbf{\cal G}(r,n)$ is constant, we can
solve this relationship for $\mbf{X}'$:

\beqno \mbf{X}'=\mbf{\cal M}(r,n) \left( \ba{c} \pp{{\cal
L}(r,n)}{\aa{r-2}{N_2-1}}\\ \pp{{\cal L}(r,n)}{\aa{r-3}{N_3-1}}\\ \vdots\\
\pp{{\cal L}(r,n)}{\aa{0}{N_r-1}} \ea \right)-\mbf{\cal M}(r,n) \left( \ba{c}
p_{1(N_2-1)}\\ p_{2(N_3-1)}\\ \vdots\\ p_{(r-1)(N_r-1)} \ea \right)-\mbf{\cal
M}(r,n)\pp{\mbf{\cal A}(r,n)}{x}, \eeqno

\no and we have expressed $\mbf{X}'$ in terms of $\mbf{q}, \mbf{p}$ and
$\mbf{X}$. Since in the previous steps, $\mbf{X}$ was expressed in terms of
$\mbf{q}$ and $\mbf{p}$, this proves that $\mbf{X}'$ is expressible in terms of
$\mbf{q}$ and $\mbf{p}$.  Continued use of the recursion relation
\rf{ostrorecur} allows us to do the same with higher derivatives of $\mbf{X}$,
which are needed as the Euler-Lagrange equations \rf{el} depend on twice
as many derivatives of the potentials as the Lagrangian. This proves that the
Ostrogradskii transformation \rf{ostrotrans} can be inverted. Hence the
Lagrangian is nonsingular if $\mbf{\cal G}(r,n)$ is nonsingular.

The converse statement is clearly also true: if the matrix $\mbf{\cal G}(r,n)$
is singular, then the Ostrogradksii transformation is not invertible (step 1 in
the proof fails, since \rf{labeliguess} cannot be solved for $\mbf{X}$). Hence
the Lagrangian is singular if the matrix $\mbf{\cal G}(r,n)$ is singular.
\hspace*{\fill}$\bbox$

\vspace*{12pt}

In sharp contrast to the Korteweg-deVries hierarchy \cite{nmpz}, the KP
hierarchy contains both singular and nonsingular Lagrangians. This is an
indication that the set of potentials $\mbf{\alpha}(r)=(\alpha_{0}(r),
\alpha_1(r), \ldots, \alpha_{r-2}(r))^T$ is not a good set of variables to
describe the dynamics of the system. These points are further explained in
Section \ref{sec:sing}.

Denote by 
\beq\la{count}
N=\sum_{i=2}^r N_i=N_2+N_3+\ldots+N_r.
\eeq

We have the following theorem:

\begin{theo}\la{ostrotheo}{\bf (Ostrogradskii \cite{dfn2, ostro})}
If the Lagrangian ${\cal L}(r,n)$ is nonsingular, then the first-order system
obtained by rewriting the Euler-Lagrange equations in terms of the new
variables $\mbf{q}$ and $\mbf{p}$ is Hamiltonian with Hamiltonian
\beq\la{ostrohamil}
H(\mbf{q},\mbf{p})=\sum_{i=1}^{r-1}\sum_{j=1}^{N_{i+1}} p_{ij} \aa{r-i-1}{j}-
{\cal L}(r,n).
\eeq
Here the inverse Ostrogradskii transformation is to be used in order to
express the Hamiltonian in terms of $\mbf{q}$ and $\mbf{p}$ only. 
The Euler-Lagrange
equations are equivalent to the $N$-dimensional Hamiltonian system
\beq\la{ostrodynsys}
\pp{q_{ij}}{x}=\pp{H}{p_{ij}},~~~\pp{p_{ij}}{x}=-\pp{H}{q_{ij}},
\eeq
for $i=1,2,\ldots, r-1$ and $j=1,2,\ldots, N_{i+1}$.
\end{theo}

\no{\bf Proof} The proof is identical to the proof in \cite{dfn2}, except that
more than one potential appears in the Lagrangian. This slight modification does
not change the proof in any fundamental way.

\vspace*{12pt}

In other words, the variables $q_{ij}$ and $p_{ij}$ are canonically conjugate
variables under the classical symplectic structure where the Poisson bracket
is given by 
\beq\la{cpb}
\{f,g\}=\sum_{i=1}^{r-1}\sum_{j=1}^{N_{i+1}}\left(\pp{f}{q_{ij}}\pp{g}{p_{ij}}-
\pp{f}{p_{ij}}\pp{g}{q_{ij}}\right)=(\mbf{\nabla} f)^T J (\mbf{\nabla} g),
\eeq

\no and 
\beqno\la{J}
\mbf{J}=\left(
\ba{cc}
\mbf{0}_N&\mbf{I}_N\\
-\mbf{I}_N & \mbf{0}_N 
\ea
\right),
\eeqno

\no where $\mbf{0}_N$ is the $N \times N$-null matrix, $\mbf{I}_N$ is the
$N\times N$-identity matrix, and
\bea\nonumber
\mbf{\nabla} f&=&\left(
\pp{f}{q_{11}},
\ldots,
\pp{f}{q_{1N_2}},
\pp{f}{q_{21}},
\ldots, 
\pp{f}{q_{2N_3}},
\ldots,
\pp{f}{q_{(r-1)N_r}}\right.,\\\la{nabla}
&&\left.\pp{f}{p_{11}},
\ldots,
\pp{f}{p_{1N_2}},
\pp{f}{p_{21}},
\ldots, 
\pp{f}{p_{2N_3}},
\ldots,
\pp{f}{p_{(r-1)N_r}}
\right)^T
\eea

\no is a $2N$-dimensional vector.

\begin{theo}
The Hamiltonian in \rf{ostrohamil} can be rewritten in the form
\bea\nonumber
H(\mbf{q},\mbf{p})&=&
\frac{1}{2} \left(\mbf{P}-\mbf{{\cal A}}(r,n)(\mbf{q})\right)^T 
\mbf{{\cal M}}(r,n)
\left(\mbf{P}-\mbf{{\cal A}}(r,n)(\mbf{q})\right)
+\\\la{goodform}
&&\sum_{i=1}^{r-1}\sum_{j=1}^{N_{i+1}-1}
p_{ij} q_{i(j+1)}-\tilde{\cal L}(r,n)(\mbf{q}),
\eea
\no with $\mbf{P}=(p_{1N_2}, p_{2N_3}, \ldots, p_{(r-1)N_r})^T$.
\end{theo}

\no{\bf Proof} (using \rf{ostrotrans}, \rf{quadraticform} and \rf{xintermsofp})
\begin{eqnarray*}
H(\mbf{q},\mbf{p})&\hspace*{-4pt}=\hspace*{-4pt}&\sum_{i=1}^{r-1}\sum_{j=1}^{N_{i+1}} p_{ij} \aa{r-i-1}{j}-
{\cal L}(r,n)\\
&\hspace*{-4pt}=\hspace*{-4pt}&\sum_{i=1}^{r-1}\sum_{j=1}^{N_{i+1}-1}p_{ij} 
\aa{r-i-1}{j}+\sum_{i=1}^{r-1}
p_{iN_{i+1}}\aa{r-i-1}{N_{i+1}}-
\frac{1}{2}\mbf{X}^T \mbf{\cal G}(r,n) \mbf{X}-\mbf{\cal A}^T(r,n)(\mbf{q})
\mbf{X}-\\&&\tilde{\cal L}(r,n)(\mbf{q})\\
&\hspace*{-4pt}=\hspace*{-4pt}&\sum_{i=1}^{r-1}\sum_{j=1}^{N_{i+1}-1}p_{ij} q_{i(j+1)}+
\mbf{P}^T \mbf{X}
-\frac{1}{2}\mbf{X}^T \mbf{\cal G}(r,n) \mbf{X}-
\mbf{\cal A}^T(r,n)(\mbf{q}) \mbf{X}-\tilde{\cal L}(r,n)(\mbf{q})\\ 
&\hspace*{-4pt}=\hspace*{-4pt}&\sum_{i=1}^{r-1}\sum_{j=1}^{N_{i+1}-1}p_{ij} q_{i(j+1)}+
\mbf{P}^T \mbf{\cal M}(r,n)
\left(\mbf{P}-\mbf{\cal A}(r,n)(\mbf{q})\right)-\\
&&\frac{1}{2}\left(\mbf{P}-\mbf{\cal A}(r,n)(\mbf{q})\right)^T 
\mbf{\cal M}^T(r,n)
\mbf{\cal G}(r,n)\mbf{\cal M}(r,n) \left(\mbf{P}-\mbf{\cal A}(r,n)(\mbf{q})
\right)-\\
&&\mbf{\cal A}^T(r,n)(\mbf{q}) \mbf{\cal M}(r,n) \left(\mbf{P}-
\mbf{\cal A}(r,n)(\mbf{q})\right)-
\tilde{\cal L}(r,n)(\mbf{q})\\
&\hspace*{-4pt}=\hspace*{-4pt}&\sum_{i=1}^{r-1}\sum_{j=1}^{N_{i+1}-1}p_{ij} q_{i(j+1)}+
\frac{1}{2}\left(\mbf{P}-\mbf{\cal A}(r,n)(\mbf{q})\right)^T 
\mbf{\cal M}(r,n) \left(\mbf{P}-\mbf{\cal A}(r,n)(\mbf{q})\right)-\tilde{\cal
L}(r,n)(\mbf{q})\\
&\hspace*{-4pt}=\hspace*{-4pt}&\frac{1}{2} \left(\mbf{P}-\mbf{\cal A}(r,n)(\mbf{q})\right)^T 
\mbf{\cal M}(r,n)
\left(\mbf{P}-\mbf{\cal A}(r,n)(\mbf{q})\right)
+\sum_{i=1}^{r-1}\sum_{j=1}^{N_{i+1}-1}
p_{ij} q_{i(j+1)}-\tilde{\cal L}(r,n)(\mbf{q}).
\end{eqnarray*}\hspace*{\fill}$\bbox$

\vspace*{12pt}

\no Note that if the middle term were missing from \rf{goodform}, the
Hamiltonian would be of the form: Kinetic Energy plus Potential Energy. Such
Hamiltonians are  called natural \cite{arnold}. (Note that $\tilde{\cal
L}(r,n)$ plays the role of minus the potential energy by its definition
\rf{quadraticform}.) However, the term natural is conventionally only used if
the mass tensor $\mbf{{\cal M}}(r,n)$ is positive definite. This is not the
case here, as the examples will illustrate.

The next theorem is well known \cite{dfn2} for a Lagrangian
depending on one potential. 

\begin{theo}\la{prop2}
\beq\la{compare}
\pp{H}{x}=-\mbf{\alpha}'^T(r) \dd{{\cal L}(r,n)}{\mbf{\alpha}(r)}. 
\eeq
\end{theo}

\no{\bf Proof}
The proof is by direct calculation. 
\begin{eqnarray*}
\pp{H}{x}&=&\pp{}{x}\left(
\sum_{i=1}^{r-1}\sum_{j=1}^{N_{i+1}} p_{ij} \aa{r-i-1}{j}-{\cal L}(r,n)
\right)\\
&=&\sum_{i=1}^{r-1}\sum_{j=1}^{N_{i+1}}\left[
\pp{p_{ij}}{x}\aa{r-i-1}{j}+p_{ij}\pp{\aa{r-i-1}{j}}{x}
\right]-\pp{{\cal L}(r,n)}{x}\\
&=&\sum_{i=1}^{r-1}\pp{p_{i1}}{x}\pp{\alpha_{r-i-1}(r)}{x}+
\sum_{i=1}^{r-1}\sum_{j=2}^{N_{i+1}}\pp{p_{ij}}{x}\aa{r-i-1}{j}+\\&&
\sum_{i=1}^{r-1}\sum_{j=1}^{N_{i+1}}p_{ij} \aa{r-i-1}{j+1}-
\sum_{i=1}^{r-1}\sum_{j=0}^{N_{i+1}}\pp{{\cal L}(r,n)}{\aa{r-i-1}{j}}
\aa{r-i-1}{j+1}\\
&=&\sum_{i=1}^{r-1}\pp{p_{i1}}{x}\pp{\alpha_{r-i-1}(r)}{x}+
\sum_{i=1}^{r-1}\sum_{j=2}^{N_{i+1}}\pp{p_{ij}}{x}\aa{r-i-1}{j}+
\sum_{i=1}^{r-1}\sum_{j=1}^{N_{i+1}}p_{ij} \aa{r-i-1}{j+1}-\\&&
\sum_{i=1}^{r-1}\sum_{j=1}^{N_{i+1}}\pp{{\cal L}(r,n)}{\aa{r-i-1}{j}}
\aa{r-i-1}{j+1}-\sum_{i=1}^{r-1}\pp{{\cal L}(r,n)}{\alpha_{r-i-1}(r)} 
\pp{\alpha_{r-i-1}(r)}{x}\\
&\since{ostrorecur}&\sum_{i=1}^{r-1}\left[
\pp{p_{i1}}{x}-\pp{{\cal L}(r,n)}{\alpha_{r-i-1}(r)}
\right] \pp{\alpha_{r-i-1}(r)}{x}+
\sum_{i=1}^{r-1}\sum_{j=2}^{N_{i+1}}\left[
\pp{{\cal L}(r,n)}{\aa{r-i-1}{j-1}}-p_{i(j-1)}
\right] \aa{r-i-1}{j}+\\
&&\sum_{i=1}^{r-1}\sum_{j=1}^{N_{i+1}}p_{ij} \aa{r-i-1}{j+1}-
\sum_{i=1}^{r-1}\sum_{j=1}^{N_{i+1}}\pp{{\cal L}(r,n)}{\aa{r-i-1}{j}}
\aa{r-i-1}{j+1}\\
&\since{ostrotrans}&-\sum_{i=1}^{r-1} \left[
\pp{{\cal L}(r,n)}{\alpha_{r-i-1}(r)}-\pp{}{x}\dd{{\cal
L}(r,n)}{\alpha_{r-i-1}'(r)}
\right]\pp{\alpha_{r-i-1}(r)}{x}+\sum_{i=1}^{r-1}\sum_{j=1}^{N_{i+1}-1}
\pp{{\cal L}(r,n)}{\aa{r-i-1}{j}} \aa{r-i-1}{j+1}
-\\&&\sum_{i=1}^{r-1}\sum_{j=1}^{N_{i+1}-1}p_{ij}\aa{r-i-1}{j+1}+
\sum_{i=1}^{r-1}
\sum_{j=1}^{N_{i+1}-1} p_{ij}
\aa{r-i-1}{j+1}+\sum_{i=1}^{r-1}p_{iN_{i+1}}\aa{r-i-1}{N_{i+1}+1}-\\
&&\sum_{i=1}^{r-1}\sum_{j=1}^{N_{i+1}-1}\pp{{\cal L}(r,n)}{\aa{r-i-1}{j}}
\aa{r-i-1}{j+1}-
\sum_{i=1}^{r-1}\pp{{\cal L}(r,n)}{\aa{r-i-1}{N_{i+1}}}
\aa{r-i-1}{N_{i+1}+1}\\
&=&-\sum_{i=1}^{r-1}\dd{{\cal L}(r,n)}{\alpha_{r-i-1}(r)}
\pp{\alpha_{r-i-1}(r)}{x}+\sum_{i=1}^{r-1}\left[
p_{iN_{i+1}}-\pp{{\cal L}(r,n)}{\aa{r-i-1}{N_{i+1}}}
\right]\aa{r-i-1}{N_{i+1}+1}\\
&=&-\mbf{\alpha}'^T(r) \dd{{\cal L}(r,n)}{\mbf{\alpha}(r)}.
\end{eqnarray*}\hspace*{\fill}$\bbox$

The simplest consequence of Theorem \ref{prop2} is that the Hamiltonian is
conserved along trajectories of the system ($i.e.$, where the Euler-Lagrange
equations \rf{el} are satisfied). But this result is a direct
consequence of Ostrogradskii's theorem: since the resulting canonical
Hamiltonian system is autonomous, the Hamiltonian is conserved. However,
Theorem \ref{prop2} will be useful when we construct more conserved
quantities in the next section. It shows how the Hamiltonian fits in
with the other conserved quantities.

\section{Complete integrability of the Hamiltonian system}\la{sec:comp}

Denote 
\beq
S(r,n)=\{H(r,k)| k=1,2, \ldots, \mbox{with} ~k~\mbox{not an integer multiple
of}~r~\mbox{or}~n\}.
\eeq

We know that in the quotient space where the Poisson bracket \rf{pb} is
defined, using \rf{pbcommute} we have
\beq\la{commute}
\left\{H(r,k_1), H(r,k_2)\right\}=\left(\dd{H(r,k_1)}{\mbf{\alpha}(r)}\right)^T
\mbf{J}(r)
\left(\dd{H(r,k_2)}{\mbf{\alpha}(r)}\right)=0.
\eeq

\no In particular, in the quotient space ($i.e.$, up to total derivatives)
\beq\la{commute2}
\left\{{\cal L}(r,n),H(r,k)\right\}=0,
\eeq

\no for $H(r,k)\in S(r,n)$. In other words, the Poisson bracket
of these two quantities is a total derivative,  
\beq\la{ostroconsalmost}
\left\{{\cal L}(r,n),H(r,k)\right\}=\pp{T_k}{x}. 
\eeq

Along the trajectories of the Euler-Lagrange equations, $i.e.$, along the
trajectories of the Hamiltonian system \rf{ostrodynsys}, the quantities $T_k$
are conserved.  A list of conserved quantities for the system \rf{ostrodynsys}
is hence obtained from 
\beq\la{ostrocons}
T_k(\mbf{q},\mbf{p})=\int \left(\dd{{\cal L}(r,n)}{\mbf{\alpha}(r)}\right)^T
\mbf{J}(r)
\left(\dd{H(r,k)}{\mbf{\alpha}(r)}\right) dx,
\eeq

\no where $k$ is not an integer multiple of $r$ or $n$. The inverse
Ostrogradskii transformation has to be used to express the right-hand side of
this equation in terms of $\mbf{q}$ and $\mbf{p}$. Note that \rf{ostrocons} is
analogous to the expression for the conserved quantities in
\cite{bogoyavlenskii}.

From the previous section, we know the Hamiltonian \rf{ostrohamil} is a
conserved quantity for the system \rf{ostrodynsys}. The theorem below shows
that up to a sign, the Hamiltonian is the first member of the newly
constructed list of conserved quantities, \rf{ostrocons}.

\begin{theo}
\beq\la{cons1}
H(\mbf{q},\mbf{p})=-T_1(\mbf{q},\mbf{p})
\eeq
\end{theo}

\no{\bf Proof} From Theorem \ref{prop2}, 
\beqno
\pp{H}{x}=-\mbf{\alpha}'^T(r) \dd{{\cal L}(r,n)}{\mbf{\alpha}(r)}=
-\left(\dd{{\cal L}(r,n)}{\mbf{\alpha}(r)}\right)^T \pp{\mbf{\alpha}(r)}{x}.
\eeqno

But the first flow of each hierarchy is the $x$-flow, therefore
\begin{eqnarray*}
\pp{H}{x}&=&-\left(\dd{{\cal L}(r,n)}{\mbf{\alpha}(r)}\right)^T \mbf{J}(r) 
\pp{H(r,1)}{\mbf{\alpha}(r)}\\
&=&-\pp{T_1}{x},
\end{eqnarray*}

\no from which we obtain the theorem. \hspace*{\fill}$\bbox$

\vspace*{12pt}

The $N$-dimensional Hamiltonian system \rf{ostrodynsys} and a list of conserved
quantities \rf{ostrocons} for it have been constructed. Complete integrability
in the sense of Liouville \cite{arnold} can be concluded under the following
conditions:

\begin{itemize}

\item In the phase space spanned by the independent coordinates $\mbf{q}$ and
$\mbf{p}$ there are $N$ nontrivial functionally independent conserved
quantities. 

\item These $N$ conserved quantities are mutually in involution with respect
to the Poisson bracket \rf{cpb}, $i.e.$, their mutual Poisson brackets vanish. 

\end{itemize}  

The list of conserved quantities \rf{ostrocons} for $k$ not an integer
multiple of $r$ or $n$ is infinite. It is clearly impossible for all of these
conserved quantities to be functionally independent: a dynamical system in a 
$2 N$-dimensional phase space has at most $2N$ independent conserved
quantities. In particular, a $2N$-dimensional autonomous Hamiltonian system
has at most $N$ independent integrals of the motion that are mutually in
involution. Below it is shown that all the conserved quantities
$T_k(\mbf{q},\mbf{p})$ are mutually in involution; therefore at most $N$ of
them are functionally independent.  We wait until Theorem
\ref{theo:atthispoint} to show that exactly $N$ different
$T_k(\mbf{q},\mbf{p})$ are nontrivial and functionally independent.

In the rest of this section, it is proved that all the $T_{k}(\mbf{q},\mbf{p})$
are in involution. This is not done by direct inspection of the mutual Poisson
brackets. Instead, we follow the approach of Bogoyavlenskii and Novikov
\cite{bogoyavlenskii}: we know from Adler \cite{adler} that all flows of the
hierarchy \rf{hamsys} commute. If we denote by $X_{t_k}$ the vector field that
evolves the phase space variables in the direction $t_k$, then the fact that
all the flows in \rf{hamsys} commute can be equivalently stated as the mutual
commutation of the vector fields $X_{t_k}$. Consider the Hamiltonian system
with canonical variables $(\mbf{q},\mbf{p})$, Poisson bracket \rf{cpb} and
Hamiltonian $H_k(\mbf{q},\mbf{p})=-T_k(\mbf{q},\mbf{p})$. We show below that
the vector field of this system, $X_{H_k}$, is the restriction of $X_{t_k}$ to
the phase space $(\mbf{q},\mbf{p})$ of the finite-dimensional Hamiltonian
system. So, the different $t_k$-flows commute, even when they are restricted to
the phase space consisting of the $(\mbf{q},\mbf{p})$ variables.  In particular
the $t_k$- and the $t_1$-flow ($i.e.$, the $x$-flow) commute. Hence we have a
family of mutually commuting Hamiltonian systems, all with the same Poisson
bracket \rf{cpb}. In \cite{arnold}, it is proved that then the family of
Hamiltonians have mutually vanishing Poisson brackets. As a consequence, the
$T_k(\mbf{q},\mbf{p})$ are mutually in involution and the system
\rf{ostrodynsys} is completely integrable in the sense of Liouville, which is
what we set out to prove.

We remark however, that this way of proving complete integrability also
provides us with $N$-dimensional Hamiltonian systems for the evolution of the
phase space variables, not only in $x$, but in all `time' variables $t_k$.
These different Hamiltonian systems have a common list of conserved quantities
in involution. Hence each is completely integrable in the sense of
Liouville. This will be spelt out in more detail once we finish proving that
all the $T_k(\mbf{q}, \mbf{p})$ are in involution.

The following lemma is used in the proof of the Bogoyavlenskii-Novikov theorem.

\begin{lemma}\la{lemmalemma}
\bea\la{lemma1}
\frac{\p^2 H}{\p p_{ij} \p q_{i(j+1)}}&=&1, ~~~~~\mbox{for}~j\leq
N_{i+1}-1\\\la{lemma2}
\frac{\p^2 H}{\p p_{ij} \p q_{is}}&=&0, ~~~~~\mbox{for}~s\neq j+1, j\leq
N_{i+1}-1\\\la{lemma3}
\frac{\p^2 H}{\p p_{i_1 j_1} \p q_{i_2 j_2}}&=&0,~~~~~\mbox{for}~i_1\neq i_2
\eea
\end{lemma}

\no{\bf Proof} We use the form \rf{goodform} of the Hamiltonian. We get 
\begin{eqnarray*}
\pp{H}{p_{ij}}=q_{i(j+1)},~~~~~\mbox{for}~j\leq N_{i+1}-1, 
\end{eqnarray*}
from which \rf{lemma1} and \rf{lemma2} easily follow. Also, \rf{lemma3}
follows from this if $j_1\leq N_{i_1+1}-1$ and ${j_2 \leq~N_{i_2+1}-1}$.
For other values of $j_1, j_2$, \rf{lemma3} follows immediately
from \rf{goodform}.\hspace*{\fill}$\bbox$

The following theorem generalizes the fundamental idea of Bogoyavlenskii and 
Novikov \cite{bogoyavlenskii}:

\begin{theo}{\bf (Bogoyavlenskii-Novikov)}
On the solution space of the $(r,n)$-th stationary KP equation, the action of
the $k$-th time variable $t_k$ can be written as an $N$-dimensional Hamiltonian
system with Hamiltonian $H_k=-T_k$ and the same canonical variables as
determined by Ostrogradksii's theorem for the $(r,n)$-th stationary KP 
equation. 
\end{theo}

\newpage

\no{\bf Proof} The proof consists of four steps:

\been

\item Prove that for $i=1, 2, \ldots, r-1$
\beq\la{step1}
\pp{q_{i1}}{t_k}=\pp{H_k}{p_{i1}}.
\eeq

\item This is an induction step. Assuming the validity of step 1, prove that
for $i=1,2,\ldots, r-1$ and $j=1,2,\ldots, N_{i+1}-1$
\beq\la{step2}
\pp{q_{i(j+1)}}{t_k}=\pp{H_k}{p_{i(j+1)}}.
\eeq

\item Prove that for $i=1, 2, \ldots, r-1$
\beq\la{step3}
\pp{p_{iN_{i+1}}}{t_k}=-\pp{H_k}{q_{iN_{i+1}}}.
\eeq

\item The last step is a `backwards' induction step: assuming step 3 is valid,
show that for $i=1,2,\ldots, r-1$ and $j=N_{i+1}+1, \ldots, 2,1$
\beq\la{step4}
\pp{p_{i(j-1)}}{t_k}=-\pp{H_k}{q_{i(j-1)}}.
\eeq

\een

\no{\bf Proof of step 1:}

During the course of this step, the index $i$ can attain any value from
$1,2,\ldots, r-1$.

Using the definition of the variational derivative,
\begin{eqnarray*}
\dd{{\cal L}(r,n)}{\alpha_{r-i-1}(r)}&=&\pp{{\cal
L}(r,n)}{\alpha_{r-i-1}(r)}-\frac{\p}{\p x} \dd{{\cal
L}(r,n)}{\alpha_{r-i-1}'(r)}\\
&=&\pp{{\cal
L}(r,n)}{\alpha_{r-i-1}(r)}-\pp{p_{i1}}{x}.
\end{eqnarray*}

\no This shows that the Euler-Lagrange equations are essentially the equations
of motion for the variables $p_{i1}, ~i=1,2,\ldots, r-1$. The other equations
of motion in Ostrogradskii's theorem are merely a consequence of the way the
new variables in the Ostrogradskii transformation are introduced.

On the other hand, from the definition of the Hamiltonian \rf{ostrohamil}, 
\begin{eqnarray*}
\pp{H}{q_{i1}}&=&-\pp{{\cal L}(r,n)}{q_{i1}}\\
&=&-\pp{{\cal L}(r,n)}{\alpha_{r-i-1}(r)}.
\end{eqnarray*}

Combining the two results, 
\beq\la{step11}
\dd{{\cal L}(r,n)}{\alpha_{r-i-1}(r)}=-\pp{H}{q_{i1}}-\pp{p_{i1}}{x}.
\eeq

We expand $\p T_k/\p x$ in two different ways:
\begin{eqnarray*}
\pp{T_k}{x}&=&\left(\dd{{\cal L}(r,n)}{\mbf{\alpha}(r)}\right)^T \mbf{J}(r) 
\left(\dd{H(r,k)}{\mbf{\alpha}(r)}\right)\\
&=&\sum_{i=1}^{r-1}\sum_{j=1}^{r-1}\dd{{\cal L}(r,n)}{\alpha_{i-1}(r)} 
J_{ij}(r)\dd{H(r,k)}{\alpha_{j-1}(r)}\\
&\since{step11}&-\sum_{i=1}^{r-1}\sum_{j=1}^{r-1}\left(\pp{H}{q_{(r-i)1}}+
\pp{p_{(r-i)1}}{x}\right)J_{ij}(r)\dd{H(r,k)}{\alpha_{j-1}(r)}, 
\end{eqnarray*}

\no and
\begin{eqnarray*}
\pp{T_k}{x}&=&\sum_{i=1}^{r-1} \sum_{j=1}^{N_{i+1}}\left(\pp{T_k}{q_{ij}}
\pp{q_{ij}}{x}+\pp{T_k}{p_{ij}}\pp{p_{ij}}{x}\right).
\end{eqnarray*}

As long as we do not impose the Euler-Lagrange equations, the derivatives of
the phase space variables are independent variations. Hence their coefficients
in both expressions for $\p T_k/\p x$ are equal. Expressing this equality for
the coefficient of $\p p_{i1}/ \p x$ gives
\begin{eqnarray*}
\pp{T_k}{p_{i1}}&=&-\sum_{j=1}^{r-1}\left(J_{(r-i)j}(r) 
\dd{H(r,k)}{\alpha_{j-1}(r)}\right)\\
&\since{hamsys}&-\pp{\alpha_{r-i-1}(r)}{t_k}\\
&\since{ostrotrans}&-\pp{q_{i1}}{t_k},
\end{eqnarray*}

\no or 
\beqno
\pp{q_{i1}}{t_k}=\pp{H_k}{p_{i1}},
\eeqno

\no which we needed to prove.\hspace*{\fill}$\bbox$ 

\no{\bf Proof of step 2:}

Assume 
\beqno
\pp{q_{i\tilde{j}}}{t_k}=\pp{H_k}{p_{i\tilde{j}}},
\eeqno

\no for $\tilde{j}=1,2,\ldots, j$. Then
\begin{eqnarray*}
\pp{q_{i(j+1)}}{t_k}&\since{ostrotrans}&\pp{\aa{r-i-1}{j}}{t_k}\\
&\since{ostrotrans}&\pp{}{x}\pp{q_{ij}}{t_k}\\
&=&\pp{}{x}\pp{H_k}{p_{ij}},
\end{eqnarray*}

\no since the $x$-flow and the $t_k$-flow commute. In the last step the
induction hypothesis is used. For any function of the variable $x$,
\bea\nonumber
\pp{f}{x}&=&\sum_{i=1}^{r-1}\sum_{j=1}^{N_{i+1}}\left(\pp{f}{q_{ij}}
\pp{q_{ij}}{x}+\pp{f}{p_{ij}}\pp{p_{ij}}{x}\right)\\\nonumber
&=&\sum_{i=1}^{r-1}\sum_{j=1}^{N_{i+1}}\left(\pp{f}{q_{ij}}
\pp{H}{p_{ij}}-\pp{f}{p_{ij}}\pp{H}{q_{ij}}\right)\\\la{step21}
&=&\{f,H\},
\eea
 
\no by definition of the Poisson bracket. With this well-known result, the above
becomes
\begin{eqnarray*}
\pp{q_{i(j+1)}}{t_k}&=&\left\{\pp{H_k}{p_{ij}}, H\right\}\\
&\since{cpb}&\left\{\left\{q_{ij},H_k\right\},H\right\}\\
&=&-\left\{\left\{H_k,H\right\},q_{ij}\right\}-
\left\{\left\{H,q_{ij}\right\},H_k\right\}\\
&\since{step21}&-\left\{\pp{H_k}{x},q_{ij}\right\}+
\left\{\left\{q_{ij},H\right\},H_k\right\}\\
&=&\left\{\left\{q_{ij},H\right\},H_k\right\}\\
&\since{step21}&\left\{\pp{q_{ij}}{x},H_k\right\}\\
&\since{ostrotrans}&\left\{q_{i(j+1)},H_k\right\}\\
&\since{cpb}&\pp{H_k}{p_{i(j+1)}},
\end{eqnarray*}

\no where we have used the fact that the Poisson bracket \rf{cpb} satisfies the
Jacobi identity. This is what we needed to prove.\hspace*{\fill}$\bbox$
 
\no{\bf Proof of step 3:}

From the commutativity of the $x$-flow and the $t_k$-flow,
\begin{eqnarray}\nonumber
\pp{}{x}\pp{q_{iN_{i+1}}}{t_k}&=&\pp{}{t_k}\pp{q_{iN_{i+1}}}{x}\\\la{step31}
\Rightarrow~~~~~~~~\pp{}{x}\pp{H_k}{p_{iN_{i+1}}}&=&\pp{}{t_k}
\pp{H}{p_{iN_{i+1}}},
\end{eqnarray}
 
\no using steps 1 and 2 of the proof. We examine the left-hand side of this
equation separately.
\begin{eqnarray*}
\mbox{Left-hand side}&=&\pp{}{x} \pp{H_k}{p_{iN_{i+1}}}\\
&\since{step21}&\left\{\pp{H_k}{p_{}iN_{i+1}},H\right\}\\
&\since{cpb}&-\left\{\left\{H_k,q_{iN_{i+1}}\right\},H\right\}\\
&=&\left\{\left\{q_{iN_{i+1}},H\right\},H_k\right\}+
\left\{\left\{H, H_k\right\},q_{iN_{i+1}}\right\}\\
&\since{step21}&\left\{\pp{q_{iN_{i+1}}}{x},H_k\right\}-\left\{
\pp{H_k}{x},q_{iN_{i+1}}\right\}\\
&=&\left\{\pp{H}{p_{iN_{i+1}}},H_k\right\},
\end{eqnarray*}

\no again using the Jacobi identity. The factor $\p H/\p p_{iN_{i+1}}$ now
appears both on the left and on the right of our equation. From \rf{goodform}
\begin{eqnarray*}
\hspace*{-12pt}&&\pp{H}{p_{iN_{i+1}}}\\
\hspace*{-12pt}&&=\pp{}{p_{iN_{i+1}}}\left(\frac{1}{2} \left(\mbf{P}-\mbf{\cal
A}(r,n)(\mbf{q})\right)^T \mbf{\cal M}(r,n) \left(\mbf{P}-\mbf{\cal
A}(r,n)(\mbf{q})\right)
\right)\\
\hspace*{-12pt}&&=\pp{}{p_{iN_{i+1}}}\left(\frac{1}{2}\sum_{i=1}^{r-1} \sum_{j=1}^{r-1} 
{\cal M}_{ij}(r,n) \left(p_{iN_{i+1}}-{\cal A}_i(r,n)(\mbf{q})\right) 
 \left(p_{jN_{j+1}}-{\cal A}_j(r,n)(\mbf{q})\right)\right)\\
\hspace*{-12pt}&&=\sum_{j=1}^{r-1}{\cal M}_{ij}(r,n) \left(p_{jN_{j+1}}-{\cal
A}_j(r,n)(\mbf{q})
\right).
\end{eqnarray*}

Using this result in \rf{step31}, 
\begin{eqnarray*}
&&\pp{}{t_k}\sum_{j=1}^{r-1} {\cal M}_{ij}(r,n)\left(
p_{jN_{j+1}}-{\cal A}_j(r,n)(\mbf{q})
\right)\\
&&=\left\{\sum_{j=1}^{r-1} {\cal M}_{ij}(r,n)
\left(p_{jN_{j+1}}-{\cal A}_j(r,n)(\mbf{q})\right), H_{k}\right\}\\
&&=\sum_{j=1}^{r-1} {\cal M}_{ij}(r,n)\left\{\left(p_{jN_{j+1}}
-{\cal A}_j(r,n)(\mbf{q})\right), H_{k}\right\}\\
&&=-\sum_{j=1}^{r-1} {\cal M}_{ij}(r,n)\left(\pp{H_k}{q_{jN_{j+1}}}+
\left\{{\cal A}_j(r,n)(\mbf{q}), H_k\right\}
\right).
\end{eqnarray*}

\no Multiplying both sides of this equation by ${\cal G}_{si}(r,n)$ and 
summing over $i$ from $1$ to $r-1$, this becomes
\begin{eqnarray*}
\pp{p_{sN_{s+1}}}{t_k}-\pp{{\cal A}_s(r,n)(\mbf{q}),}{t_k}=
-\pp{H_k}{q_{sN_{s+1}}}-\left\{{\cal A}_s(r,n)(\mbf{q}), H_k\right\}.
\end{eqnarray*}

Since ${\cal A}_s(r,n)(\mbf{q})$ depends only on $\mbf{q}$, it follows from
step 2 that the second term on the left-hand side is equal to the second term
on the right-hand side. Hence
\beqno
\pp{p_{sN_{s+1}}}{t_k}=-\pp{H_k}{q_{sN_{s+1}}},
\eeqno

\no for $s=1,2,\ldots, r-1$. This
terminates the proof of step 3. \hspace*{\fill}$\bbox$

Note that it is necessary for the Lagrangian ${\cal L}(r,n)$ to be nonsingular,
as we are using the matrix ${\cal M}(r,n)$.

\no{\bf Proof of step 4:}

Assume 
\beqno
\pp{p_{i\tilde{j}}}{t_k}=-\pp{H_k}{q_{i\tilde{j}}},
\eeqno

\no for $\tilde{j}=N_{i+1}, N_{i+1}-1, \ldots, j$. We have 
\begin{eqnarray*}
\pp{}{t_k}\pp{p_{ij}}{x}&=&\pp{}{x}\pp{p_{ij}}{t_k}\\
&\since{step21}&\left\{\pp{p_{ij}}{t_k},H\right\}\\
&=&-\left\{\pp{H_k}{q_{ij}},H\right\}\\
&\since{cpb}&\left\{\left\{p_{ij},H_k\right\},H\right\}\\
&=&-\left\{\left\{H_k, H\right\},p_{ij}\right\}-
\left\{\left\{H, p_{ij}\right\},H_k\right\}\\
&\since{step21}&-\left\{\pp{H_k}{x}, H\right\}+\left\{\pp{p_{ij}}{x},
H_k\right\}\\
&=&-\left\{\pp{H}{q_{ij}}, H_k\right\}\\
&\since{cpb}&
-\sum_{\gamma=1}^{r-1}\sum_{\delta=1}^{N_{\gamma+1}}\left( \frac{\p^2 H}{\p
q_{ij}\p q_{\gamma \delta}}\pp{H_k}{p_{\gamma \delta}}-\frac{\p^2 H}{\p q_{ij}
\p p_{\gamma \delta}}\pp{H_k}{q_{\gamma\delta}}\right)\\
&=&-\sum_{\gamma=1}^{r-1}\sum_{\delta=1}^{N_{\gamma+1}} \frac{\p^2 H}{\p
q_{ij}\p q_{\gamma \delta}}\pp{H_k}{p_{\gamma \delta}}+
\sum_{\gamma=1}^{r-1}\sum_{\delta=1}^{N_{\gamma+1}}\frac{\p^2 H}{\p q_{ij}
\p p_{\gamma \delta}}\pp{H_k}{q_{\gamma\delta}}\\
&\since{lemma3}&-\sum_{\gamma=1}^{r-1}\sum_{\delta=1}^{N_{\gamma+1}} 
\frac{\p^2 H}{\p
q_{ij}\p q_{\gamma \delta}}\pp{H_k}{p_{\gamma \delta}}+\sum_{\delta=1}^{N_{i+1}}
\frac{\p^2 H}{\p q_{ij} \p p_{i \delta}} \pp{H_k}{q_{i\delta}}\\
&\since{lemma1}&-\sum_{\gamma=1}^{r-1}\sum_{\delta=1}^{N_{\gamma+1}} 
\frac{\p^2 H}{\p
q_{ij}\p q_{\gamma \delta}}\pp{H_k}{p_{\gamma \delta}}+\frac{\p^2 H}{\p q_{ij}
\p p_{i(j-1)}} \pp{H_k}{q_{i(j-1)}}+\\&&
\frac{\p^2 H}{\p q_{ij} \p p_{iN_{i+1}}}
\pp{H_k}{q_{iN_{i+1}}}\\
&\since{lemma2}&-\sum_{\gamma=1}^{r-1}\sum_{\delta=1}^{N_{\gamma+1}} 
\frac{\p^2 H}{\p
q_{ij}\p q_{\gamma \delta}}\pp{H_k}{p_{\gamma \delta}}+\pp{H_k}{q_{i(j-1)}}.
\end{eqnarray*}

The left-hand side of this equation can be expressed another way as well:
\begin{eqnarray*}
\pp{}{t_k}\pp{p_{ij}}{x}&=&\pp{}{t_k}\left(-\pp{H}{q_{ij}}\right)\\
&=&-\sum_{\gamma=1}^{r-1}\sum_{\delta=1}^{N_{\gamma+1}}\left(\frac{\p^2 H}{\p
q_{ij} \p q_{\gamma \delta}}\pp{q_{\gamma \delta}}{t_k}+\frac{\p^2 H}{\p q_{ij}
\p p_{\gamma \delta}} \pp{p_{\gamma \delta}}{t_k}\right)\\
&=&-\sum_{\gamma=1}^{r-1}\sum_{\delta=1}^{N_{\gamma+1}}\left(\frac{\p^2 H}{\p
q_{ij} \p q_{\gamma \delta}}\pp{H_k}{p_{\gamma \delta}}
+\frac{\p^2 H}{\p q_{ij}
\p p_{\gamma \delta}} \pp{p_{\gamma \delta}}{t_k}\right)\\
&=&-\sum_{\gamma=1}^{r-1}\sum_{\delta=1}^{N_{\gamma+1}}\frac{\p^2 H}{\p
q_{ij} \p q_{\gamma \delta}}\pp{H_k}{p_{\gamma \delta}}-\pp{p_{i(j-1)}}{t_k},
\end{eqnarray*}

\no where the second term has been simplified using lemma \ref{lemmalemma}, as
before. Comparing the two right-hand sides, the double-summed term drops out
and one finds
\beqno
\pp{p_{i(j-1)}}{t_k}=-\pp{H_k}{q_{i(j-1)}}.
\eeqno 

This completes the proof of step 4 and hence of the Bogoyavlenskii-Novikov
theorem. \hspace*{\fill}$\bbox$

\vspace*{12pt}

Let us recapitulate the most important results of the last few sections. 

\begin{itemize}

\item The $(r,n)$-th stationary KP equation can be written as an $N$-dimensional
Hamiltonian system in $x$, given by Ostrogradskii's theorem \rf{ostrodynsys}.

\item This Hamiltonian system is completely integrable in the sense of
Liouville. $N$ independent conserved quantities in involution
$T_k(\mbf{q},\mbf{p})$ can be
constructed explicitly.

\item These conserved quantities can be interpreted as Hamiltonians: The
$t_k$-flow induces on the solution space of the $(r,n)$-th KP equation an
evolution which is Hamiltonian with Hamiltonian $H_k=-T_k$. This Hamiltonian
system shares its phase space variables, symplectic structure and conserved
quantities with the $x$-system. The $t_k$-evolution of the phase space
variables is given by
\beq\la{tkdynsys}
\pp{q_{ij}}{t_k}=\pp{H_k}{p_{ij}},~~~~\pp{p_{ij}}{t_k}=-\pp{H_k}{q_{ij}},
\eeq

\no for $i=1,2,\ldots, r-1$ and $j=1,2,\ldots, N_{i+1}$. The Hamiltonian is
given by
\beq\la{tkham}
H_k(\mbf{q},\mbf{p})=-T_{k}(\mbf{q},\mbf{p}).
\eeq

This only gives non-trivial results only if $k$ is not a multiple of $r$ and
$n$. Strictly speaking however, this is not required for the proof of the
Bogoyavlenskii-Novikov theorem.

\end{itemize}

At this point, we have established enough results
to argue that $N$ of the conserved quantities $T_k(r,n)$ are nontrivial and
functionally independent:

\begin{theo}\la{theo:atthispoint}
In the phase space spanned by the independent coordinates $\mbf{q}$ and
$\mbf{p}$ there are $N$ nontrivial functionally independent conserved
quantities
\end{theo}

\no{\bf Proof} 
In the previous theorem, we have shown that the conserved
quantity $T_k(\mbf{q},\mbf{p})$ is minus the Hamiltonian for the evolution  of
the phase space variables $(\mbf{q},\mbf{p})$ in the $t_k$-direction. So, if
$T_k(\mbf{q},\mbf{p})$ is trivial ($i.e.$, $T_k(\mbf{q},\mbf{p})$=0 on the
entire phase space), then so is the dependence of the solution of the
$(r,n)$-th KP equation, and conversely. A typical solution of the $(r,n)$-th
KP equation is a solution of genus $N$ of the KP equation (see \cite{ds1}) of
the form 

\beq\la{compkpsol} u=u_0+2\p_x^2 \ln \Theta(\sum_{j=1}^\infty
\mbf{K}_j t_j).  
\eeq 

\no Here all the $\mbf{K}_j$ are $N$-dimensional vectors.  If the conserved
quantity $T_k(\mbf{q}, \mbf{p})$ is functionally dependent on any of the other
$T_j(\mbf{q}, \mbf{p})$, $j<k$, then the vectorfield $X_{H_k}$ is a linear
combination of the vectorfields $X_{H_j}, j<k$. Hence the vector $\mbf{K}_k$
is a linear combination of the vectors $\mbf{K}_j, j<k$.  If $\mbf{K}_k$, is
linearly dependent on the vectors with lower indices, we can use a linear
transformation to obtain a solution of the form \rf{compkpsol} which depends
on $t_1, t_2, \ldots, t_{k-1}$, but is independent of $t_k$ (for instance,
this is possible for $t_r$ and $t_n$). If $\mbf{K}_k$ is independent of the
vectors $\mbf{K}_j$, with $j<l$, then the solution depends on $t_k$ in a
nontrivial manner. In this case, the conserved quantity $T_k(\mbf{q},\mbf{p})$
has to be nontrivial and functionally independent of $T_j$, for $j<k$. A
linear space of dimension $N$ is spanned by $N$ linearly independent vectors.
Hence a typical solution of the $(r,n)$-th KP equation has $N$ nontrivial
functional independent conserved quantities $T_k(\mbf{q},\mbf{p})$.
\hspace*{\fill}$\bbox$

\vspace*{12pt}

\no{\bf Remark}
One often convenient way to integrate an integrable Hamiltonian
system explicitly is analogous to the method of forward and inverse scattering,
but restricted to systems of ordinary differential equations \cite{wojo}. To
invoke this method, a Lax representation \cite{wojo} for the system of
Hamiltonian equations is required. Such a representation was obtained in step 5
of the algorithm presented in \cite{ds1}. 

\vspace*{12pt}

In what follows, only the Hamiltonian system in $x$ is considered. Any
conclusions reached are however also valid for the Hamiltonian systems in any
of the `time' variables $t_k$.

\section{Examples}

In this section, the abstract formalism of the previous sections is
illustrated using concrete examples, by assigning concrete values to $r$ and
$n$. The simplest cases are discussed: $r=2$ and various values for $n$ (the
KdV hierarchy); $r=3$ and various values for $n$ (the Boussinesq hierarchy). A
special case of $r=3$ gives rise to stationary three-phase solutions of the KP
equation, namely for $n=4$. This case is illustrated in more detail than the
other examples. It is the easiest example not covered by the work of
Bogoyavlenskii and Novikov \cite{bogoyavlenskii}. 

\vspace*{12pt}

\no {\bf (a) The KdV hierarchy: one-dimensional solutions of the KP equation}
\vspace*{12pt}

The KdV hierarchy is obtained from the KP hierarchy by imposing the reduction
$r=2$, hence 
\beq\la{kdvrred}
L^2_+=L^2.
\eeq

\no Since $L^2_+=\p^2+2 u_2$, $u_2$ is the only independent potential with the
other potentials determined in terms of it. From $L^2_-=0$
\beq\la{triangkdv}
u_3=-\frac{u_2'}{2},~u_4=\frac{u_2''}{4}-\frac{u_2^2}{2}, ~u_5=\frac{3}{2} u_2
u_2' -\frac{u_2'''}{8}, ~etc.
\eeq

\no Other ingredients needed for \rf{explicitlaxrred} are: $M(2)=M_{2,1}=\p$
and $\beta(2,n)=\beta_{-1}(n)=\alpha_{-1}(n)$. Hence the KdV hierarchy has the
form  \beq\la{kdvhiernonham} \pp{u_2}{t_n}=\pp{}{x} \alpha_{-1}(n). \eeq 

\no In order to write this in Hamiltonian form, we use the potential
$\alpha_0(2)=2 u_2=u$. Then $D(2)=2$, by \rf{jacobian}. Hence the Poisson
structure of the KdV hierarchy is given by $J(2)=D(2)M(2)=2\p$. Using
\rf{manin2} with $j=1$ and $r=2$,
\beq\la{kdvhiermanin}
\alpha_{-1}(2)=\beta_{-1}(2)=\frac{2}{2+n} \dd{\alpha_{-1}(2+n)}{u}
\eeq
we can recast the KdV hierarchy in its familiar Hamiltonian form \cite{gardner}:
\beq\la{kdvhierhamsys}
\pp{u}{t_n}=2 \pp{}{x}\dd{H(2,n)}{u},
\eeq

\no with $H(2,n)=2 \alpha_{-1}(2+n)/(2+n)$. If the factor 2 is absorbed into
the definition of the Hamiltonian, then this form of the KdV hierarchy is
identical to that introduced by Gardner \cite{gardner}. Note that immediately
all even flows ({\em i.e.,} $t_{n}=t_{2k}$, for $k$ a positive integer) are
trivial because $2+2k$ is not coprime with $r=2$, so
$H(2,n)=\alpha_{-1}(2+2k)/(1+k)\equiv 0$. We write out some nontrivial flows
explicitly:

\begin{description}

\item[(i)~~] $n=1$: $H(2,1)=u^2/4$ and
\beq
\pp{u}{t_1}=u_x,
\eeq 
as expected.

\item[(ii)~] $n=3$: $H(2,3)=u^3/8-u_x^2/16$ and 
\beq
\pp{u}{t_3}=\frac{1}{4}\left(6 u u_x+u_{xxx}\right),
\eeq
the KdV equation. 

\item[(iii)] $n=5$: $H(2,5)=5 u^4/64-5 u u_x^2/32+u_{xx}^2/64$ and 
\beq\la{kdv5}
\pp{u}{t_5}=\frac{1}{16}\left(30 u^2 u_x+20 u_x u_{xx}+10 u u_{xxx}+u_{5x}
\right),
\eeq
the 5th-order KdV equation. 

\end{description}

There is only one Casimir functional for the KdV hierarchy, namely
$H(2,-1)=2 \alpha_{-1}(1)=2 u_2=u$. It is easily verified that this is indeed a
Casimir functional for the Poisson structure $J(2)=2 \p$: $J(2)(\delta
H(2,-1)/\delta u)=2 \p (\delta u/\delta u)=2 \p (1)=0$. 

Imposing the $n$-reduction, the Lagrangian ${\cal L}(2,n)$ has the form
\beq\la{kdvlag} {\cal L}(2,n)=H(2,n)+\sum_{k=1}^{n-2}d_{k} H(2,k)+h_1 u.  \eeq
This Lagrangian was first obtained by Bogoyavlenskii and Novikov
\cite{bogoyavlenskii}. From Table 1, $N_2=[(n-1)/2]=(n-1)/2$, since $n$ is
odd. Necessarily, the Lagrangian has the form \beq\la{kdvhierlagform} {\cal
L}(2,n)=\frac{1}{2} a \left(u^{((n-1)/2)}\right)^2+\hat{\cal L}(2,n), \eeq for
some nonzero constant $a$ and $\hat{\cal L}(2,n)$ independent of
$u^{((n-1)/2)}$. The Lagrangian is always nonsingular. 

Because the case $r=2$ was covered by Bogoyavlenskii and Novikov, no examples
of the Ostrogradksii transformation and the resulting Hamiltonian system of
ordinary differential equations will be given. A typical solution of the
$(2,n)$-th KP equation, {\em i.e.,} a stationary solution of the $n$-KdV
equation, depends on $N=N_2=(n-1)/2$ phases. These solutions are also
one-dimensional since they are independent of $y=t_2$. 

\vspace*{12pt}

\no {\bf (b) The Boussinesq hierarchy: stationary solutions of the KP equation}

\vspace*{12pt}

The Boussinesq hierarchy is obtained from the KP hierarchy by imposing the 
reduction $r=3$, hence 
\beq\la{bousrred}
L^3_+=L^3.
\eeq

\no Since $L^3_+=\p^3+3 u_2\p+3 u_2'+3 u_3$, $u_2$ and $u_3$ are the only
independent potentials with the other potentials determined in terms of
these two. From $L^3_-=0$

\beq\la{triangbous}
u_4=-u_3'-u_2^2-\frac{u_2''}{3}, ~u_5=-2 u_2 u_3+2 u_2 u_2'+\frac{2}{3}
u_3''+\frac{u_2'''}{3}, ~etc.
\eeq

\no Furthermore, 
\beq
\mbf{U}(3)=\left(\ba{c}u_2\\u_3\ea\right), ~~
\mbf{\beta}(3,n)=\left(\ba{c}\beta_{-1}(n)\\\beta_{-2}(n)\ea\right), ~~
\mbf{M}(3)=\left(\ba{cc}\p & 0\\ -\p^2 & \p\ea\right),
\eeq

\no so that the Boussinesq hierarchy is 
\beq\la{bousshiernonham}
\pp{\mbf{U}(3)}{t_n}=\mbf{M}(3)\mbf{\beta}(3,n) ~\Rightarrow~\left\{
\ba{rcl}
\ds{\pp{u_2}{t_n}}&\ds{=}&\ds{\pp{\beta_{-1}(n)}{x}}\\
\ds{\pp{u_3}{t_n}}&\ds{=}&\ds{-\ppn{2}{\beta_{-1}(n)}{x}+\pp{\beta_{-2}(n)}{x}}
\ea
\right. .
\eeq

\no In order to write this in Hamiltonian form, we use the potentials
$\alpha_1(3)=3 u_2=u$ and $\alpha_{0}(3)=3 u_2'+3 u_3=v$, where $u$ and $v$
are introduced for notational simplicity. Then 

\beq
\mbf{\alpha}(3)=\left(
\ba{c}
v\\u
\ea
\right), 
~
\mbf{D}(3)=\left(
\ba{cc}
3\p & 3\\3&0
\ea
\right),~
\mbf{\beta}(3,n)=\frac{3}{3+n}\dd{}{\mbf{\alpha}(3)}\alpha_{-1}(3+n).
\eeq

\no Hence the Poisson structure of the Boussinesq hierarchy is
\beq\la{poissonbous}
J(3)=D(3)M(3)=3 \left(\ba{cc}0 & \p\\
\p & 0\ea\right).
\eeq

\no The Boussinesq hierarchy is written in Hamiltonian form as:
\beq\la{boushierhamsys}
\pp{}{t_n}\left(\ba{c}v\\u\ea\right)=3 \left(\ba{cc}0 & \p\\
\p & 0\ea\right) \left(\ba{c}
\ds{\dd{H(3,n)}{v}}\\
\ds{\dd{H(3,n)}{u}}
\ea\right)=3 \pp{}{x}
\left(
\ba{c}
\ds{\dd{H(3,n)}{u}}\\
\ds{\dd{H(3,n)}{v}}
\ea
\right)
,
\eeq

\no with $H(3,n)=3 \alpha_{-1}(3+n)/(3+n)$. Up to a factor 3, this form of the
Boussinesq hierarchy is identical to the one introduced by McKean
\cite{mckean1}. We write some flows explicitly:

\begin{description}

\item[(i)~~] $n=1$: $H(3,1)=uv/3$ and
\beq
\left\{
\ba{rcl}
\ds{v_{t_1}}&=&\ds{v_x}\\
\ds{u_{t_1}}&=&\ds{u_x}
\ea
\right.,
\eeq 
as expected.

\item[(ii)~] $n=2$: $H(3,2)=-u^3/27+u_x^2/9+v^2/3-v u_x/3$ and 
\beq
\left\{
\ba{rcl}
\ds{v_{t_2}}&=&\ds{-\frac{2}{3}u u_x-\frac{2}{3}u_{xxx}+v_{xx}}\\
\ds{u_{t_2}}&=&\ds{2 v_x-u_{xx}}
\ea
\right..
\eeq
Elimination of $v$ from these two equations gives the Boussinesq equation,
\beq\la{bouss}
u_{t_2t_2}+\frac{1}{3}u_{xxxx}+\frac{2}{3}\left(u^2\right)_{xx}=0.
\eeq

\item[(iii)] $n=4$: $H(3,4)=-u_{xx}^2/27+v_x u_{xx}/9-v_{x}^2/9+u u_x^2/9-2 u v
u_x/9-u^4/81+2 u v^2/9$ and 
\beq\la{bouss4}
\left\{
\ba{rcl}
\ds{v_{t_4}}&=&\ds{-\frac{4}{9}u^2 u_x+\frac{4}{3}v v_x-\frac{4}{3}u_x
u_{xx}-\frac{2}{3}u u_{xxx}+\frac{2}{3}u_x v_x+\frac{2}{3}u v_{xx}-\frac{2}{9}
u_{5x}+\frac{2}{3} v_{xxxx}}\\
&&\\
\ds{u_{t_4}}&=&\ds{-\frac{2}{3}u_x^2-\frac{2}{3}u u_{xx}+\frac{4}{3}v
u_x-\frac{2}{3}u_{xxxx}+\frac{2}{3}v_{xxx}}
\ea
\right..
\eeq
This is the next member of the Boussinesq hierarchy. 

\end{description}

There are two Casimir functionals for the Boussinesq hierarchy, namely
$H(3,-1)=3 \alpha_{-1}(2)/2=3 (u_{2}'+u_3)/2=v/2$ and $H(3,-2)=3
\alpha_{-1}(1)=3 u_2=u$. For convenience $u$ and $v$ are used as Casimir
functionals below. 

Imposing the $n$-reduction, the Lagrangian ${\cal L}(3,n)$ has the form
\beq\la{bouslag} 
{\cal L}(3,n)=H(3,n)+\sum_{k=1}^{n-2}d_{k} H(3,k)+h_1 u+h_2 v .  
\eeq

Theorem \ref{theo:sing}, given in the next section, shows that this Lagrangian
is always nonsingular.  A typical solution of the $(3,n)$-th KP equation, {\em
i.e.,} a stationary solution of the $n$-Boussinesq equation, depends on
$N=N_2+N_3=n-1$ phases.  These solutions are stationary solutions of the KP
equation, since they are independent of $t=t_3$. 

\vspace*{12pt}

\no {\bf (c) Stationary 3-phase solutions of the KP equation}

\vspace*{12pt}

Consider the $r=3$, $n=4$ reduction. The Lagrangian is
\bea\nonumber
{\cal L}(3,4)&=&-\frac{u_{xx}^2}{27}+\frac{v_x
u_{xx}}{9}-\frac{v_{x}^2}{9}+\frac{u u_x^2}{9}-\frac{2 u vu_x}{9}-
\frac{u^4}{81}+\frac{2 u v^2}{9}+\\\la{lag34}
&&d_2\left(-\frac{u^3}{27}+\frac{u_x^2}{9}+\frac{v^2}{3}-\frac{vu_x}{3}\right)+
d_1 \frac{uv}{3}+h_1 u+h_2 v.
\eea

The Ostrogradskii transformation \rf{ostrotrans} for this Lagrangian is
\bea\nonumber
q_{11}=u,&&p_{11}=\dd{{\cal
L}(3,4)}{u_x}=\frac{u_{xxx}}{54}+\frac{uu_x}{9}+\frac{d_2
u_x}{18}+\frac{d_1u}{6}+\frac{h_2}{2},\\\la{ostrotrans34}
q_{12}=u_x,&&p_{12}=\dd{{\cal L}}{u_{xx}}=-\frac{2
u_{xx}}{27}+\frac{v_x}{9},\\\nonumber
q_{21}=v,&&p_{21}=\dd{{\cal L}(3,4)}{v_x}=\frac{u_{xx}}{9}-\frac{2v_x}{9}.
\eea

\no In the definition of $p_{11}$, the Euler-Lagrange equations have been used
to eliminate $v_{xx}$. The Ostrogradskii transformation can be inverted:
\bea\nonumber
&u=q_{11},~~ u_x=q_{12},~~ v=q_{21},&\\\la{invostro34}
&u_{xx}=-54 p_{12}-27 p_{21},~~v_x=-27
p_{12}-18p_{21},&\\\nonumber
& u_{xxx}=54p_{11}-6 q_{11}q_{12}-3 d_2 q_{12}-9d_1q_{11}-27 h_2.&
\eea

\no Using the inverse Ostrogradskii transformation, the Hamiltonian
corresponding to the Lagrangian \rf{lag34} is
\bea\nonumber 
H&=&-27p_{12}^2-27 p_{12} p_{21}-9 p_{21}^2+p_{11}
q_{12}+\frac{q_{11}^4}{81}-\frac{q_{11} q_{12}^2}{9}-\frac{2 q_{21}^2
q_{11}}{9}+\frac{2 q_{11} q_{12}
q_{21}}{9}+\\\la{ham34}
&&d_2\left(\frac{q_{11}^3}{27}-\frac{q_{12}^2}{9}-\frac{q_{21}^2}{3}+
\frac{q_{12}q_{21}}{3}\right)-d_1 \frac{q_{11} q_{21}}{3}-h_1 q_{11}-h_2 q_{21}.
\eea

\no For simplicity the constants $d_1, d_2, h_1$ and $h_2$ are equated to zero
in the remainder of this example. Using \rf{ostrocons}, three conserved
quantities are computed for the Hamiltonian system generated by \rf{ham34}:
\bea\la{t134}
T_1&=&-H\\\nonumber
T_2&=&3 \int \left(\dd{{\cal L}(3,4)}{v}\pp{}{x}\dd{H(3,2)}{u}+
\dd{{\cal L}(3,4)}{u}\pp{}{x}\dd{H(3,2)}{v}\right) dx\\\nonumber
&=&\frac{4 q_{12}q_{11}^3}{81}-\frac{8 q_{21} q_{11}^3}{81}-2
p_{12}q_{12}q_{11}-\frac{4p_{21} q_{12} q_{11}}{3}+4 p_{12}q_{11}q_{21}+2 p_{21}
q_{21}q_{11}-\frac{q_{12}^3}{27}+\\\la{t234}
&&\frac{q_{21}^3}{27}-\frac{2q_{12}q_{21}^2}{9}+9
p_{11}p_{21}+\frac{4q_{12}^2q_{21}}{27}\\\nonumber
T_5&=&3 \int \left(\dd{{\cal L}(3,4)}{v}\pp{}{x}\dd{H(3,5)}{u}+
\dd{{\cal L}(3,4)}{u}\pp{}{x}\dd{H(3,5)}{v}\right) dx\\\nonumber
&=&-\frac{2 q_{11}^6}{729}+\frac{8p_{12}q_{11}^4}{27}+\frac{4 p_{21}
q_{11}^4}{27}-\frac{q_{12}^2 q_{11}^3}{243}+3 p_{11}p_{21}q_{21}+\frac{2q_{21}^2
q_{11}^3}{243}-\frac{2 q_{12} q_{21} q_{11}^3}{243}-9
p_{12}^2q_{11}^2-\\\nonumber
&&2
p_{21}^2q_{11}^2-9 p_{12} p_{21} q_{11}^2+\frac{p_{11} q_{12} q_{11}^2}{3}-3
p_{11}^2 q_{11}+\frac{p_{21} q_{12}^2 q_{11}}{9}+\frac{q_{21}^4}{27}+54
p_{12}^3-\frac{2 q_{12} q_{21}^3}{27}+\\\la{t534}
&&27 p_{12} p_{21}^2+\frac{4 q_{12}^2
q_{21}^2}{81}+81 p_{12}^2 p_{21}-3 p_{11}p_{12} q_{12}-3
p_{11}p_{21}q_{12}-\frac{q_{12}^3q_{21}}{81}-
\frac{2 p_{21} q_{12} q_{21} q_{11}}{9}.
\eea

\no Since $r=3$ and $n=4$, the conserved quantities $T_3$ and $T_4$ are trivial.
It is easy to check by direct computation that these three conserved quantities
are in involution. Furthermore, the dependence of the solution of the KP
equation on $t_2$ ({\em i.e.,} y) is governed by the Hamiltonian system
generated by $H_2=-T_2$. The same statement holds for $t_5$ and $H_5=-T_5$. We
will return to this example in Section \ref{sec:reductions}

\vspace*{12pt}

\section{Singular and nonsingular Lagrangians}\la{sec:sing}

We have shown that the Hamiltonian system in $x$ \rf{ostrodynsys} is
completely integrable in the sense of Liouville when Ostrogradksii's theorem
applies, $i.e.$, when the Lagrangian ${\cal L}(r,n)$ is nonsingular. This has
immediate consequences for the structure of the solutions. On a compact
component of phase space, almost all solutions are quasi-periodic with $N$
phases. These phases move linearly on an $N$-dimensional torus \cite{arnold}.
Such a torus is topologically equivalent to a compact component of
\beq\la{torus} 
\Lambda(r,n)=\left\{T_k(\mbf{q},\mbf{p})={\cal T}_k, k \in\Omega(r,n)\right\} 
\eeq

\no where $\Omega(r,n)=\{\mbox{first N values of}~k, \mbox{not integer 
multiples of $r$ or $n$}\}$. The constants ${\cal T}_k$ are
determined by the initial conditions. The torus $\Lambda(r,n)$ is shared by all
$t_k$-flows. The only difference between these different flows from this point
of view is the linear motion on the torus. This linear motion determines the
$N$ frequencies with respect to the variable $t_k$.

We know from \cite{ds1} that a typical solution of the $(r,n)$-th KP equation
has $g_{max}=(r-1)(n-1)/2$ phases.  In the proof of Theorem \ref{theo:sing},
it is shown that $N=g_{max}$ if one of $r$, $n$ is even and the other one is
odd. The case of both $r$ and $n$ even is not allowed, since $r$ and $n$ need
to be coprime. On the other hand, if both $r$ and $n$ are odd, $N=g_{max}$
only if $r=3$. Otherwise $N>g_{max}$ and the Ostrogradskii transformation
introduces more variables than are needed to span the phase space. The
transformation is then not invertible and the Lagrangian ${\cal L}(r,n)$ is
nonsingular. This situation does not occur in the work of Bogoyavlenskii and
Novikov \cite{bogoyavlenskii}, because for the KdV hierarchy $r=2$. From the
results in this section it follows that some rank 1, finite-genus solutions 
(namely $r>3$ and odd, $n>r$ and odd), give rise to interesting examples of
singular Lagrangians.

\begin{theo}\la{theo:sing} 
The Lagrangian ${\cal L}(r,n)$ is singular if and only if $r$ and
$n$ are both odd and $r>3$. For all other cases of $(r,n)$, $N=g_{max}$.
\end{theo}

\no {\bf Proof} 
First note that $\left[R/2\right]+\left[(R+1)/2\right]=R$, for all integers
$R$. Using Table \ref{table1} we can rewrite 
\bea\nonumber
N&=&\sum_{j=2}^{r}N_j\\\nonumber
&=&N_2+N_3+\sum_{j=4}^{r}N_j\\\la{rewrite}
&=&r+n-4+\sum_{j=4}^{r}\left[\frac{n+r-2 j+2}{2}\right].
\eea

In the calculations below, we use that $n$ can always be chosen to be  greater
than $r$.  The calculations are slightly different depending on whether $r$
and/or $n$ are even or odd. Since $r$ and $n$ cannot both be even, there are
three cases.

\been

\item $r$ is even and $n$ is odd. We write $r=2R$, $n=2M+1$. Use \rf{rewrite},
\begin{eqnarray*}
N&=&r+n-4+\sum_{j=4}^{2R}\left(M+R-j+1\right)\\
&=&r+n-4+2RM-3M-2R+3\\
&=&\frac{(r-1)(n-1)}{2}\\
&=&g_{max}.
\end{eqnarray*}

\item $r$ is odd and $n$ is even. We write $r=2R+1$, $n=2M$. The calculations
are similar to those in the previous case. 
\begin{eqnarray*}
N&=&r+n-4+\sum_{j=4}^{2R+1}(M+R-j+1)\\
&=&\frac{(r-1)(n-1)}{2}\\
&=&g_{max}.
\end{eqnarray*}

\item $r$ is odd and $n$ is odd. We write $r=2R+1$, $n=2M+1$. In this case, the
result is quite surprising. 
\begin{eqnarray*}
N&=&r+n-4+\sum_{j=4}^{2R+1}\left(R+M-j+2\right)\\
&=&r+n-4+\frac{(R+M-2)(R+M-1)}{2}-\frac{(M-R+1)(M-R)}{2}\\
&=&\frac{(r-1)(n-1)}{2}+\frac{r-3}{2}\\
&=&g_{max}+\frac{r-3}{2}.
\end{eqnarray*}

\no Hence, in this case, $N\neq g_{max}$.  So, if $r$ and $n$ are both odd and
$r > 3$, the dimension of the torus $\Lambda(r,n)$ in \rf{torus} is seemingly
greater than the maximal number of phases of a solution of the $(r,n)$-th KP
equation, according to \cite{ds1}. This dimension exceeds the maximal genus by
the amount of $(r-3)/2$, which is a positive integer when $r$ is odd and
greater than three. This is an indication that the assumptions necessary for
Ostrogradskii's theorem have been violated. Hence $(r,n ~\mbox{both odd},
r>3)$ is a  sufficient condition for the Lagrangian ${\cal L}(r,n)$ to be
singular. 

That this condition is necessary for ${\cal L}(r,n)$ to be singular follows
from the results of Veselov \cite{veselov}. There it is demonstrated that the
dimension of the phase space of the Euler-Lagrange equations \rf{el} is always
equal to $2 g_{max}$, which is the desired dimension of the phase space of the
Hamiltonian system \rf{hamsys}. In such a case the Lagrangian is only singular
if the Ostrogradskii transformation introduces more variables than the
dimension of the phase space \cite{krupkova}, which only happens when $r,n$ are
noth odd and $r>3$. Hence this condition is also necessary for the singularity
of the Lagrangian.
\hspace*{\fill}$\bbox$

\een

Note that in the generic case of \cite{ds1}, $r=g+1$, $n=g+2$, we always have
$g_{max}=N$, since this automatically puts one in case 1 or case 2.

\vspace*{12pt}

\no {\bf Example}

\vspace*{8pt}

The smallest odd value possible for $r$ is $r=3$. In this case, the count of
the number of phases is still right. This corresponds to the Boussinesq
hierarchy. The smallest values of $r$ and $n$ where the count of the number of
phases is wrong occurs when $(r,n)=(5,7)$. This example is discussed below. 
In this example, the Lagrangian ${\cal L}(5,7)$ expressed in the $4$ variables
$\mbf{\alpha}(5)=(\alpha_0(5),\alpha_1(5),\alpha_2(5), \alpha_3(5))^T$ is
singular. It is shown below how one deals with the singular Lagrangian case.
As it turns out, a relatively straightforward transformation reduces the
Lagrangian ${\cal L}(5,7)$ to a nonsingular Lagrangian, expressed in the
transformed variables. The Ostrogradskii transformation is applicable to this
Lagrangian and it results in a completely integrable system with $N=g_{max}$.

For simplicity, denote $u=\alpha_{3}(5)$, $v=\alpha_2(5)$, $w=\alpha_1(5)$ and
$z=\alpha_0(5)$. The Lagrangian is
\bea\nonumber
{\cal L}(5,7)&=&-\frac{2u_{xxxx}}{25}\left(z-\frac{5}{2}w_x+\frac{9}{5} v_{xx}
\right)_{xx}
+\frac{w_{xx}v_{xxxx}}{5}-\frac{7 z_x
v_{xxxx}}{25}-\frac{7 z_{xx} w_{xx}}{25}+\\\la{lag57} &&\tilde{\cal
L}(u,u_x,u_{xx}, u_{xxx},v,v_x, v_{xx},v_{xxx}, w, w_x, w_{xx}, z, z_x),
\eea

\no where all dependence on the vector $\mbf{X}=(u_{xxxx}, v_{xxxx}, w_{xxx},
z_{xx})^T$ is explicit. We have
\beq\la{ganda}
\mbf{\cal G}(5,7)=\left(\ba{cccc}
0&-18/125&1/5&-2/25\\
-18/125&0&0&0\\
1/5&0&0&0\\
-2/25&0&0&0
\ea
\right),~\mbf{\cal A}(5,7)=\left(\ba{c}
0\\w_{xx}/5-7z_x/25\\0\\-7 w_{xx}/25
\ea
\right).
\eeq

\no Clearly $\mbf{\cal G}(5,7)$ is singular.  If canonical variables $\mbf{q}$
and $\mbf{p}$ were introduced using the Ostrogradskii transformation on this
Lagrangian, this would lead to $2N=2(N_2+N_3+N_4+N_5)=26$ phase space
variables. The dimension of the phase space can be obtained from the
Euler-Lagrange equations \cite{veselov} corresponding to ${\cal L}(5,7)$ and
is 24.  Since the corank of the matrix $\mbf{\cal G}(5,7)$ is 2, it follows
from the Ostrogradskii transformation \rf{ostrotrans} that two linear
combinations between $p_{14},p_{24}, p_{33}$ and $p_{42}$ exist. These are
also linear in $\mbf{q}$ because $\mbf{\cal A}(5,7)$ is linear in $\mbf{q}$:

\beq\la{lindep}
p_{24}=-\frac{2}{5}p_{33}+\frac{7}{25}q_{33},~~~p_{42}=-\frac{18}{25}p_{33}+
\frac{1}{5}q_{33}-\frac{7}{25}q_{42}.
\eeq

\no At this point the theory of constrained Hamiltonian systems can be used
\cite{dirac}. Another possibility is to use the general methods of Krupkova
\cite{krupkova} for singular Lagrangians.  However, it is possible to
transform the potentials to a new set of potentials such that the Lagrangian
is nonsingular when expressed in the new potentials. Motivated by the form of
the Lagrangian, let

\beq\la{nonsingtrans}
\hat{u}=u,~~
\hat{v}=v,~~
\hat{w}=w,~~
\hat{z}=z-\frac{5}{2}w_x+\frac{9}{5}v_{xx}.
\eeq

\no This transformation is clearly invertible. In the new variables, the
Lagrangian is 

\bea\nonumber
{\cal L}(5,7)=
-\frac{2}{25}\hat{u}_{xxxx}\hat{z}_{xx}+\frac{1}{250}\hat{v}_{xxx}\hat{w}_{xx}-
\frac{7}{25}\hat{v}_{xxxx}\hat{z}_x-\frac{7}{25}\hat{w}_{xx}\hat{}_{xx}+\\
\la{lag57transformed}
\hat{\cal
L}(\hat{u},\hat{u}_x,\hat{u}_{xx}, \hat{u}_{xxx},\hat{v},\hat{v}_x, 
\hat{v}_{xx},\hat{v}_{xxx}, \hat{w}, \hat{w}_x, \hat{w}_{xx}, \hat{z}, 
\hat{z}_x),
\eea

\no up to total derivatives. Define a new vector $\hat{\mbf{X}}=(\hat{u}_{xxxx},
\hat{v}_{xxxx}, \hat{w}_{xx}, \hat{z}_{xx})^T$. The Lagrangian is

\beq\la{newlag}
{\cal L}(5,7)=\frac{1}{2}\hat{\mbf{X}}^T \hat{\mbf{\cal G}}(5,7)
\hat{\mbf{X}}+\hat{\mbf{\cal A}}(5,7) \hat{\mbf{X}}+\hat{\cal L}(5,7),
\eeq

\no with

\beq\la{newg}
\hat{\mbf{\cal G}}(5,7)=\left(\ba{cccc}
0&0&0&-2/25\\
0&0&1/250&0\\
0&1/250&0&-7/25\\
-2/25&0&-7/25&0
\ea
\right),
\eeq

\no which is nonsingular, hence by Theorem \ref{prop:sing} the Lagrangian is
nonsingular. The Ostrogradskii transformation acting on the transformed
Lagrangian \rf{newlag} introduces 24 canonical variables, as many variables as
the dimension of the phase space \cite{veselov}. 

\vspace*{12pt}


It is not clear if the approach of this example always works, $i.e.$, for
other values of $r$ and $n$, both odd. There is no proof and the problem of
dealing with a singular Lagrangian for values of $(r,n)$ other than $(5,7)$
describing the $(r,n)$-th KP equation may require the general methods alluded
to above.

\section{Autonomous symmetry reductions of the Hamiltonian
system}\la{sec:reductions}

As discussed in the remarks on page \pageref{remrem}, not all solutions of the
$(r,n)$-th KP equation have the same number of phases. A generic genus $g$
solution of the KP equation is usually a very non-typical solution of the
$(r,n)$-th KP equation, with $r=g+1$ and $n=g+2$.

\begin{figure}[htb]
\centerline{\psfig{file=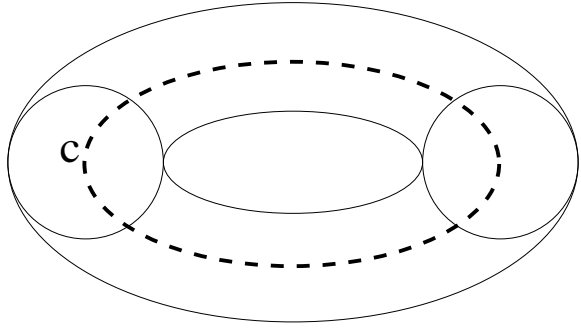,width=3in}}
\caption{\label{fig1} {\bf Usually solutions evolve on 2-D tori. Special
solutions correspond to lower-dimensional tori.}}
\end{figure}

A solution of the $(r,n)$-th KP equation is completely determined by the
phase-space variables $q$ and $p$. In the $2N$-dimensional phase space
coordinatized by $\mbf{q}$ and $\mbf{p}$, for a given initial condition, the
solution evolves on the torus $\Lambda(r,n)$ determined by \rf{torus}. Usually
($i.e.$, for almost all initial data in the class of solutions of the
$(r,n)$-th KP equation), this torus is $N$-dimensional \cite{arnold}. However,
special solutions correspond to lower-dimensional tori. For example, suppose
that $N=2$. Then almost all solutions evolve on a two-dimensional torus: for
almost all values of ${\cal T}_1$ and ${\cal T}_2$, the surface $\Lambda$
determined by $T_1={\cal T}_1$ and  $T_2={\cal T}_2$ is topologically
equivalent to a two-dimensional torus, like the one drawn in Figure
\ref{fig1}. For special values of ${\cal T}_1$ and ${\cal T}_2$, however, 
this torus is degenerate and the resulting `surface' is only a one-dimensional
torus, $i.e.$, a circle $C$. This is drawn in Figure \ref{fig1}.

To a certain degree this scenario is the typical one that occurs. For almost
all values of the constants $\{{\cal T}_k, ~k\in \Omega(r,n)\}$, the torus
$\Lambda(r,n)$ in \rf{torus} is $N$-dimensional. For a special set of values 
$\{{\cal T}_k, ~k\in \Omega(r,n)\}$, the torus is $(N-1)$-dimensional,
corresponding to a solution with $(N-1)$ phases. For an even more limited
class of constants $\{{\cal T}_k, ~k\in \Omega(r,n)\}$, the torus
$\Lambda(r,n)$ is $(N-2)$-dimensional, corresponding to  solutions with
$(N-2)$ phases, $etc$. 

The conditions on the set of constants $\{{\cal T}_k, ~k\in \Omega(r,n)\}$ can
be expressed in terms of the variables $\mbf{q}$ and $\mbf{p}$. When these
variables satisfy certain constraints (to be derived below), the dimension of
the torus $\Lambda(r,n)$ decreases.

We have argued above that the conserved quantities $T_k$ for $k \in
\Omega(r,n)$ are functionally independent quantities if the $\mbf{q}$ and
$\mbf{p}$ variables are considered independent variables. The only way for the
torus $\Lambda(r,n)$ to be less than $N$-dimensional is if the conserved
quantities are $not$ functionally independent. This is only possible if there
are functional relationships between the variables $(\mbf{q},\mbf{p})$.
The constraints on the variables $\mbf{q}$ and $\mbf{p}$ are obtained as
follows:

\been

\item Require that the conserved quantities $T_k$ for $k \in \Omega(r,n)$ be
functionally dependent. This is expressed as 
\beq\la{funcdep}
\mbox{rank}\left(
\nabla T_{i_1}~\nabla T_{i_2}~\ldots~\nabla T_{i_N} 
\right)=g<N,
\eeq

where the indices $i_k$ are the elements of $\Omega(r,n)$ and $\nabla T_{i_k}$
is defined in \rf{nabla}. In this case, $N-g$
of the conserved quantities are functionally dependent on the remaining $g$
functionally independent conserved quantities. Without loss of generality, we
assume that the first $g$ conserved quantities remain functionally independent,
whereas the last $N-g$ conserved quantities are functionally dependent:
\beq\la{firstg}
T_{i_k}=F_{i_k}(T_{i_1}, T_{i_2}, \ldots, T_{i_g}),
\eeq

\no for $k=g+1, g+2, \ldots, N$.

\item In this case, there are only $g$ functionally independent conserved
quantities $T_{i_k}$, $k=1,2,\ldots, g$. The manifold \rf{torus} reduces to 
\beq\la{reducedtorus}
\Lambda_g(r,n)=\{T_{i_k}={\cal T}_{k}, ~k=1,2,\ldots,g\},
\eeq

which is topologically equivalent to a $g$-dimensional torus. This
$g$-dimensional torus is parametrizable by $g$ phase variables moving
linearly on the torus. In other words, if the evolution of a solution of the
$(r,n)$-th KP equation is restricted to this $g$-dimensional torus, it has
only $g$ phases. Since this solution is also a solution of the KP equation and
it has rank 1 and $g$ phases, it must be a genus $g$, rank 1 solution of the
KP equation.

\item Equations \rf{funcdep} results in a number of conditions on the 
variables $q$ and $p$:
\beq\la{pqcon}
K_j(\mbf{q},\mbf{p})=0, ~~~~j=1, 2, \ldots, m,
\eeq

\no where $m$ is the number of conditions. If these conditions are satisfied
for the `initial' conditions $\mbf{q}(0)$ and $\mbf{p}(0)$ then they are
automatically satisfied for all other $x$-values, $i.e.$, the conditions on the
variables $\mbf{q}$ and $\mbf{p}$ are invariant under the $x$-flow. This is
easy to see: The conditions \rf{pqcon} are equivalent to the conditions
\rf{funcdep} which are equivalent to the conditions \rf{firstg}, which only
involve the conserved quantities. Clearly \rf{firstg} are invariant conditions.

The conditions on the variables $\mbf{q}$ and $\mbf{p}$ \rf{pqcon} are
polynomial in the $\mbf{q}$ and $\mbf{p}$ variables, since all the entries of
the matrix on the left-hand side of \rf{funcdep} are polynomial in these
variables. In practice, the conditions on the variables $\mbf{q}$ and $\mbf{p}$
\rf{funcdep} can be written as combinations of simpler conditions,

\beq\la{simplercon}
K_j=\sum_{k=1}^{m_g} Q_{j,k}(\mbf{q},\mbf{p}) P_k(\mbf{q},\mbf{p}), 
~~~~j=1, 2, \ldots, m.
\eeq

\no Here both $Q_{j,k}(\mbf{q},\mbf{p})$ and $P_k(\mbf{q},\mbf{p})$ are
polynomials in $q$ and $p$. If $P_{k}(\mbf{q},\mbf{p})=0$, for $k=1,2, \ldots,
m_g$ then clearly the conditions \rf{pqcon} are satisfied. Clearly the
decomposition \rf{simplercon} is not unique. The factors $P_k$ of a
given decomposition are not necessarily invariant under the $x$-flow. In order
to find a minimal ($i.e.$, smallest number of elements) set of conditions on
the $\mbf{q}$ and $\mbf{p}$ variables, the invariance of such factors needs to
be tested separately. Since the conditions \rf{funcdep} are invariant, as
argued above, such a minimal set of invariant factors is guaranteed to exist.
The existence of a {\em minimal} decomposition is essentially a restatement
of Hilbert's Basis theorem \cite{abhyankar}. Below, we prove that the number
of elements in this set is $m_g=2(N-g)$. Once this minimal set of conditions
has been found, \rf{pqcon} can be replaced by the conditions
\beq\la{pqmincon}
P_k(\mbf{q},\mbf{p})=0, ~~~k=1,2,\ldots, m_g.
\eeq

\no The invariance of the factors $P_k(\mbf{q},\mbf{p})$ is easily tested. It
is necessary and sufficient that \cite{olver}
\beq\la{inv}
\left\{P_k(\mbf{q},\mbf{p}),H\right\}=0,~~~~~~\mbox{for~} k=1,2,\ldots, m_g,
\eeq

\no on the solution manifold $P_k(\mbf{q},\mbf{p})=0$, for $k=1,2,\ldots, m_g$.

\item The conditions on the variables $\mbf{q}$ and $\mbf{p}$ \rf{pqmincon} are
autonomous, since the conditions do not depend explicitly on $x$. The
conditions \rf{pqmincon} determine {\em autonomous invariant symmetry
reductions} of the Hamiltonian system \rf{ostrodynsys}.

\een

\begin{theo}\cite{dullin}\la{dull}
In order for a solution of the $(r,n)$-th KP equation to have genus
$g$ instead of $N$, $2(N-g)$ conditions need to be imposed on the variables
$\mbf{q}$
and $\mbf{p}$, $i.e.$, $m_{g}=2(N-g)$. 
\end{theo}

\no{\bf Proof} By Ostrogradskii's theorem, a typical solution of the $(r,n)$-th
KP equation resides in the $2N$-dimensional phase space with coordinates
$\mbf{q}$
and $\mbf{p}$. The existence of $N$ conserved quantities $T_{i_k}$, for
$k=1,2,\ldots, N$ guarantees that the motion starting from any initial
condition evolves on a torus determined by the $N$ conditions $T_{i_k}={\cal
T}_k$, for $k=1,2,\ldots, N$. Hence this torus is a hypersurface of codimension
$N$, or of dimension $2N-N=N$.

If we impose that the rank of the matrix $\left(\nabla T_{i_1}, \nabla T_{i_2},
\ldots, \nabla T_{i_N}\right)$ is $N-1$ in order to obtain genus $N-1$
solutions, then the motion of the solutions is restricted to an
$(N-1)$-dimensional torus. An $(N-1)$-dimensional hypersurface in a
$2N$-dimensional phase space is determined by $N+1$ equations. $N-1$ equations
are provided by the relations $T_{i_k}={\cal T}_k$, for $k=1,2,\ldots, N-1$.
Hence another two conditions are required on the coordinates $\mbf{q}$ and
$\mbf{p}$. 

The proof of the theorem is now easily obtained by repeating this argument $N-g$
times. \hspace*{\fill}$\bbox$

\vspace*{12pt}

\no {\bf Remarks}

\begin{description}

\item[(a)~]
The fact that $m_{N-1}=2$ is not easily seen from \rf{funcdep}. For
\rf{funcdep} to be satisfied with $g=N-1$, the determinants of all $N\times N$
minors need to be zero. This seemingly results in $\binomial{2N}{N}$
$N$-dimensional minors of which $2N-1$ are functionally independent. The
determinant of all these minors are decomposable in terms of two polynomials
$P_1(\mbf{q},\mbf{p})$ and $P_2(\mbf{q},\mbf{p})$, which are both invariant
under the $x$-flow. This is explicitly worked out in the example below, for
$N=3$ and $g=2$. 

\item[(b)~] It should be mentioned that in \cite{ds1}, some ideas were given
to find conditions on nontypical solutions of the $(r,n)$-th KP equation.
Those ideas were correct, but seemingly hard to implement, as is seen in the
example discussed there. This same example is discussed below. The algorithm
offerered in this section to find the determining conditions on the nontypical
solutions of the $(r,n)$-th KP equation is much more efficient.

\end{description}

\no {\bf Example: Generic genus 2 solutions of the KP equation}

\vspace*{6pt}
 
\no A generic solution of genus 2 of the KP equation is a solution of the
$(3,4)$-th KP equation. The Hamiltonian system for this case was discussed on
page \pageref{ham34}. The Hamiltonian system \rf{ostrodynsys} corresponding to
the Hamiltonian \rf{ham34} is three-dimensional, hence a typical solution of
this system executes linear motion on a 3-torus, topologically equivalent to
\beq\la{3torus} \Lambda(3,4)=\left\{T_1={\cal T}_1,~T_2={\cal T}_2, ~T_5={\cal
T}_5 ~\right\}. \eeq

\no Such a solution has three phases and is hence a rank 1, genus 3 solution
of the KP equation. Nevertheless, the generic rank 1, genus 2 solutions of the
KP equation are obtained as solutions of the $(3,4)$-th KP equation. These
solutions are special, nontypical solutions of the $(3,4)$-th KP equation.
They are obtained by imposing an autonomous invariant symmetry reduction on
the Hamiltonian system \rf{ostrodynsys}, as outlined in this section. 

The condition \rf{funcdep} for a solution of the $(3,4)$-th KP equation to
have genus 2 is
\beq\la{g2con}
\mbox{rank}\left(
\ba{cccccc}
\ds{\pp{T_1}{q_{11}}} & \ds{\pp{T_1}{q_{12}}} & \ds{\pp{T_1}{q_{21}}} &
\ds{\pp{T_1}{p_{11}}} & \ds{\pp{T_1}{p_{12}}} & \ds{\pp{T_1}{p_{21}}}\\
\vspace{-5pt}\\
\ds{\pp{T_2}{q_{11}}} & \ds{\pp{T_2}{q_{12}}} & \ds{\pp{T_2}{q_{21}}}&
\ds{\pp{T_2}{p_{11}}} & \ds{\pp{T_2}{p_{12}}} & \ds{\pp{T_2}{p_{21}}}\\
\vspace{-5pt}\\
\ds{\pp{T_3}{q_{11}}} & \ds{\pp{T_3}{q_{12}}} & \ds{\pp{T_3}{q_{21}}}&
\ds{\pp{T_3}{p_{11}}} & \ds{\pp{T_3}{p_{12}}} & \ds{\pp{T_3}{p_{21}}}\\
\ea
\right)=2.
\eeq

From Theorem \ref{dull}, it follows that \rf{g2con} is equivalent to two
invariant conditions on the variables $q_{11}$, $q_{12}$, $q_{21}$, $p_{11}$,
$p_{12}$ and $p_{21}$. These conditions are readily found in this specific
case \cite{decthesis}. The expressions for these conditions are quite long,
so we do not repeat them here. Let us denote them by 
\beq\la{g2cond}
P_1(\mbf{q}, \mbf{p})=0,~~p_2(\mbf{q}, \mbf{p})=0.
\eeq

There is a more geometrical way to look at the conditions \rf{g2con}. Consider
the three-dimensional space spanned by the conserved quantities $T_1, T_2$ and
$T_5$. If the conditions \rf{g2cond} are satisfied, there is a functional
relationship between the three conserved quantities: \beq\la{funcrelcons}
\Omega: f(T_1,T_2,T_5)=0, \eeq

\no which represents a surface in the space spanned by $T_1, T_2$ and $T_5$.
By solving the conditions \rf{g2cond} for two of the phase space variables
($p_{11}$ and $p_{12}$ respectively, for instance), and substituting the
result in the form of the conserved quantities \rf{t134}, \rf{t234} and
\rf{t534}, a parametric representation of the surface $\Omega$ is obtained:
\beq\la{g2parametric} \Omega:\left\{ \ba{rcl} T_1&=&T_1(q_{11}, q_{12}, q_{21},
p_{21}),\\ T_2&=&T_2(q_{11}, q_{12}, q_{21}, p_{21}),\\ T_5&=&T_3(q_{11},
q_{12}, q_{21}, p_{21}). \ea \right. \eeq

\begin{figure}
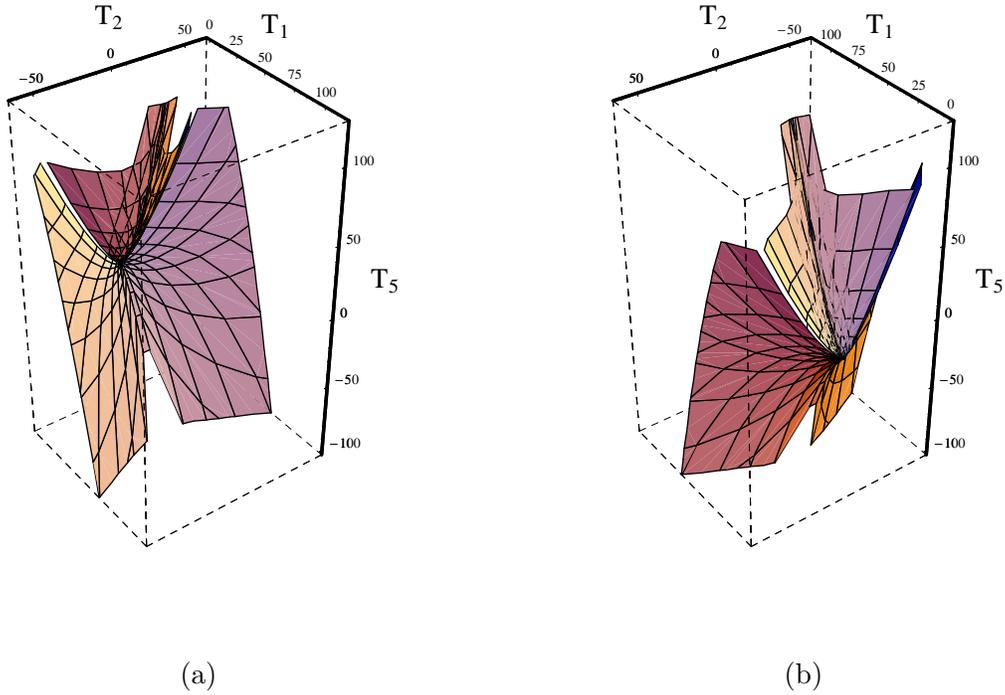

\begin{tabular}{cc}
\psfig{file=surf1.ps,width=3in} & 
\psfig{file=surf2.ps,width=3in}\\
(a) & (b)\\
\end{tabular}
\caption{\label{fig2} {\bf The surface $\Omega$ from two different view points.
The cusp of the surface is at the origin. Figure (b) shows the same surface as 
Figure (a), but rotated around the $T_5$-axis by 180 degrees. This figure was
obtained using \rf{g2parametric} with $q_{21}=0=p_{21}$.}}
\end{figure}

\no Apparently, two too many parameters are present in this set of equations,
since the parametric representation of a surface should only contain two
parameters. However, the existence of a functional relationship
\rf{funcrelcons} guarantees that these parameters appear only in two different
combinations such that there are essentially two parameters present in
\rf{g2parametric}. The most convenient way to plot the resulting surface is to
equate two of the `parameters' $q_{11}$, $q_{12}$, $q_{21}$ and $p_{21}$ to
zero, while letting the remaining two vary. In Figure \ref{fig2}, two
different views of the surface $\Omega$ are given.

Every point in the space spanned by $T_1$, $T_2$ and $T_5$ corresponds to a
three-dimensional torus, on which the motion of the solutions of the $(3,4)$-th
KP equation takes place. A point on the surface $\Omega$ corresponds to a
degenerate three-dimensional torus, on which there are in essence only two
independent directions, analogous to the idea demonstrated in Figure
\ref{fig1}. In other words, points on the surface $\Omega$ correspond to
two-dimensional tori and to genus two solutions of the KP equation.
These genus two solutions are the generic rank 1, genus 2 solutions of the KP
equation, as argued above.

Note that a more drastic reduction to genus one solutions is possible. These
solutions correspond to points on the singular curves on the surface $\Omega$.
As the genus one solutions are more easily obtained from the $r=2$ case, this
case is not essential.

\section{Parameter count}\la{sec:parameters}

The previous sections put us in a position to count easily the
parameters that determine a solution of the $(r,n)$-th KP equation. 

The first column of table \ref{table2} lists the parameters that determine a
solution of the $(r,n)$-th KP equation. They are the `initial' values variables
$\mbf{q}(0)$ and $\mbf{p}(0)$ (not only for the $x$-flow, but any other
$t_k$-flow), the constants $h_k$ that are the coefficients of the Casimir
functionals in \rf{lagrangian}, the constants $d_k$ which are the coefficients
of the lower-order flows in \rf{lagrangian} and the constant $u_1$, discussed 
in the remark on page \pageref{u1rem}.

\settowidth{\mylength}{$\ds{N=\frac{(r-1)(n-1)}{2}}$}
\settoheight{\myheight}{\framebox{$\ds{N=\frac{(r-1)(n-1)}{2}}$}}
\addtolength{\myheight}{8pt}

\begin{table}
\begin{center}
\caption{\bf The number of parameters determining the solutions of the 
$(r,n)$-th KP
equation.\la{table2}}
\vspace*{0.2in}
\begin{tabular}{|c|c|c|c|}
\hline
& typical solution of & generic genus $g$ & non-typical solution of \\
&the $(r,n)$-th KP equation& solution of (KP)& the $(r,n)$-th KP equation\\
\hline\hline
\pb{$\mbf{q}(0)$} & $\ds{N=\frac{(r-1)(n-1)}{2}}$ & $g$ & $g$\\
\hline
\pb{$\mbf{p}(0)$} & $\ds{N=\frac{(r-1)(n-1)}{2}}$ & $g$ & $g$\\
\hline\hline
\pb{$h_k$} & $r-1$ & $g$ & $r-1$\\
\hline
\pb{$d_k$} & $n-2$ & $g$ & $n-2$\\
\hline
\pb{$u_1$} & 1 & 1 & 1\\
\hline\hline
\pb{total \#} & $rn-1$ & $4g+1$ & $2g+r+n-2$\\
\hline
\end{tabular}
\end{center}
\end{table}

How many of each of these parameters determine a typical solution of the
$(r,n)$-th KP equation is indicated in the second column. A typical solution of
the $(r,n)$-th KP equation has $N=(r-1)(n-1)/2$ variables $\mbf{q}$ and  $N$
variables $\mbf{p}$. Each of these is determined by its initial conditions
$\mbf{q}(0)$  and $\mbf{p}(0)$. For a typical solution of the $(r,n)$-th KP
equation, $N$ is also the genus of the solution. Any solution of the $(r,n)$-th
KP equation is determined by $r-1$ Casimir functionals (see \rf{lagrangian}).
Also from \rf{lagrangian}, it follows that $n-2$ lower-order flows are
included, accounting for the coefficients $d_k$. With the addition of the
constant $u_1$, this results in a total number of $rn-1$ parameters that
determine a typical solution of the $(r,n)$-th KP equation.

The third column expresses the number of parameters in terms of the genus of
the solution, for a generic genus $g$ solution of the KP equation. For such a
solution $r=g+1$ and $n=g+2$ \cite{ds1}. Furthermore, the
Hamiltonian system \rf{ostrodynsys} reduces significantly such that there are
exactly $g$ variables $\mbf{q}$ and $\mbf{p}$. The total number of parameters
adds up to $4g+1$. This is a known result, implied for instance in \cite{dub}.

Not every nontypical solution of the $(r,n)$-th KP equation is generic. For
such solutions, the number of variables $\mbf{q}$ is equal to the genus $g$ of
the solution, as is the number of variables $\mbf{p}$. These results are given
in the last column of Table \ref{table2}.

There is an important distinction between the different types of parameters in
table \ref{table2}. The entries in the top two rows have dynamical
significance: they are initial values for the variables $\mbf{q}$ and
$\mbf{p}$. The Hamiltonian system \rf{ostrodynsys} is a dynamical system for
the determination of the variables $\mbf{q}$ and $\mbf{p}$. The other
parameters merely show up as parameters in this Hamiltonian system. This
distinction between two kinds of parameters, dynamical and nondynamical,
is to our knowledge new. 

\section{Minimal Characterization of the initial data}

A rank 1, finite-genus solution of the KP equation is completely determined by
a solution of an $(r,n)$-th KP equation, for a certain $r$ and $n$. The
$(r,n)$-th KP equation is given by the Euler-Lagrange equation \rf{el}. This
is a set of $(r-1)$ ordinary differential equations in $x$. Various quantities
appear in this system of ordinary differential equations: $(r-1)$ potentials
$\mbf{\alpha}(r)$ and their derivatives with respect to $x$, $(r-1)$ constants
$h_k$, $(n-2)$ constants $d_k$ and one constant potential $u_1$. 

Next we argue that the knowledge of the initial condition for KP, $u(x,y,t=0)$
along one direction (say the $x$-axis) is sufficient to determine the
corresponding rank 1, finite genus solution for all $x$, $y$ and $t$. 

\been

\item If the initial condition is specified along the $x$-axis for $y=0$, all
potentials and their derivatives can be found at any point on the $x$-axis
(Note that a rank 1, finite-genus solution is analytic in all its independent
variables). This is done in the following way: The Euler-Lagrange equations
and their derivatives with respect to $x$ determine $algebraic$ conditions on
the potentials and their derivatives, as well as the unknown parameters $h_k$,
$d_k$ and $u_1$. In these conditions, $u$ and its derivatives with respect to
$x$ are known, by assumption. Hence taking more derivatives of the
Euler-Lagrange equations \rf{el} with respect to $x$ adds more conditions than
unknowns. Taking enough derivatives of the Euler-Lagrange equations, we
obtain a set of polynomial equations for the unknown potentials and their
derivatives, as well as the parameters $h_k$, $d_k$ and $u_1$.  

\item We have already argued that the knowledge of the $x$-dependence
completely determines the dependence of the solution on any higher-order time
variable: the Hamiltonians determining the dependence of the solution on $t_k$
are conserved quantities for the $x$-evolution of the solution, $H_k=-T_k$.
Furthermore, the initial conditions $\mbf{q}(0)$, $\mbf{p}(0)$ for $t_k$ are
identical to the initial conditions for the $x$-evolution at $x=0$, $y=0$.

\een

This shows that it suffices to specify the rank 1, finite-genus initial
condition along one direction: $u(x, y=0, t=0)$. It is clear that the above
argument can be repeated if the initial condition is specified along any of the
higher-order flows. This is specifically relevant for $t_2~(y)$ and $t_3~(t)$.

\vspace*{12pt}

\no {\bf Remarks} The procedure for finding the parameters $h_k$, $d_k$ and
$u_1$ given above is not very practical. A large number of derivatives of the
potential are required and a large polynomial system needs to be solved to
find the values of the parameters. 

\section*{Acknowledgements} The author acknowledges useful discussions with O.
I. Bogoyavlenskii and A. P. Veselov. H. Segur is thanked for his continued
support and the invaluable interactions with the author on this subject. 
This work was carried out at the University of Colorado and
supported by NSF grant DMS-9731097.

\bibliographystyle{unsrt}

\bibliography{}

\begin{center}

Mathematical Sciences Research Institute\\
1000 Centennial Drive\\
Berkeley Ca 94720\\
Tel: (510) 643-6022\\
Fax: (510) 642-8609\\
e-mail:deconinc@msri.org

\end{center}

\end{document}